\documentclass[preprint,12pt,authoryear]{elsarticle}

\usepackage{amssymb}

\usepackage{mathrsfs}
\usepackage{enumerate}
\usepackage{amsmath,amssymb}
\usepackage{graphicx,epsfig,natbib}
\usepackage{bibentry}
\usepackage{appendix}
\usepackage{hyperref}
\usepackage{caption,subcaption,placeins,float}
\usepackage{tikz,sidecap}
\usepackage{bbm,comment,url}
\usepackage{mathtools}
\usepackage{rotating,multirow}
\usepackage[margin=1in]{geometry}

\usepackage[utf8]{inputenc}
\usepackage{tcolorbox}
\usepackage{graphicx}
\usepackage{amsmath}
\usepackage{amssymb}
\usepackage{amsfonts}
\usepackage{natbib}
\usepackage{bibentry}
\usepackage{hyperref}
\usepackage{caption}
\usepackage{subcaption}

\usepackage{algpseudocode}
\usepackage{algorithm}

\usepackage{xcolor}

\usepackage{bm}
\usepackage{mathrsfs}

\newcommand{\mr}{\mathrm}

\newcommand{\mc}{\mathcal}
\newcommand{\ms}{\mathsf}
\newcommand{\bs}{\boldsymbol}

\newcommand{\y}{\bs{y}}

\newcommand{\D}{\mc{D}_{\mathrm{sim}}}

\newcommand{\bftheta}{\bs{\theta}}

\newcommand{\Ndata}{N}
\newcommand{\m}{\ms{m}}

\newcommand{\inVec}{\boldsymbol{\theta}}
\newcommand{\outVec}{\boldsymbol{y}}
\newcommand{\outSca}{y}
\newcommand{\bfu}{{\bf u}}

\newcommand{\Cij}{K}
\newcommand{\COmatrix}{{\bf A}}
\newcommand{\COij}{A}

\renewcommand{\vec}[1]{\bm{\mathrm{#1}}}

\def \CC{\mathbb{C}}

\def \FF{\mathbb{F}}

\def \T{\vec{T}}

\def \fhg{\vec{f}}

\def \half{\frac{1}{2}}

\usepackage{multirow}
\usepackage{prettyref}
\newrefformat{sec}{Section~\ref{#1}}
\newrefformat{subsec}{Section~\ref{#1}}
\newrefformat{fig}{Figure~\ref{#1}}
\newrefformat{app}{Appendix~\ref{#1}}
\newrefformat{tab}{Table~\ref{#1}}
\newrefformat{alg}{Algorithm~\ref{#1}}

\global\long\def\mr#1{\mathrm{#1}}

\global\long\def\ms#1{\mathsf{#1}}

\global\long\def\mc#1{\mathcal{#1}}

\global\long\def\bs#1{\boldsymbol{#1}}

\global\long\def\T{\top}

\global\long\def\D{\mc D}

\global\long\def\m{\mathsf{m}}

\journal{Journal of the Royal Statistical Society, Series C}

\begin{document}

\begin{frontmatter}



\title{{\bf Fast Parameter Inference in a Biomechanical Model of the Left Ventricle using Statistical Emulation}}


\author[label1,label2]{Vinny Davies\fnref{fn1}}
\author[label3]{Umberto No\`{e}\fnref{fn1}}
\author[label2]{Alan Lazarus}
\author[label2]{Hao Gao}
\author[label2]{Benn Macdonald}
\author[label4,label5]{Colin Berry}
\author[label2]{Xiaoyu Luo}
\author[label2]{Dirk Husmeier\corref{cor1}}
\cortext[cor1]{Corresponding author.}
\ead{Dirk.Husmeier@glasgow.ac.uk}

\fntext[fn1]{Authors contributed equally.}

\address[label1]{School of Computing Science, University of Glasgow, Glasgow, UK.}
\address[label2]{School of Mathematics and Statistics, University of Glasgow, Glasgow, UK.}
\address[label3]{German Centre for Neurodegenerative Diseases (DZNE), Bonn, Germany.}
\address[label4]{BHF Glasgow Cardiovascular Research Centre, University of Glasgow, Glasgow, UK.}
\address[label5]{West of Scotland Heart and Lung Centre, Golden Jubilee National Hospital, Clydebank, UK.}

\begin{abstract}
A central problem in biomechanical studies of personalised human left ventricular (LV) modelling is estimating the material properties and biophysical parameters from in-vivo clinical measurements in a time frame suitable for use within a clinic. Understanding these properties can provide insight into heart function or dysfunction and help inform personalised medicine. However, finding a solution to the differential equations which mathematically describe the kinematics and dynamics of the myocardium through numerical integration can be computationally expensive. To circumvent this issue, we use the concept of emulation to infer the myocardial properties of a healthy volunteer in a viable clinical time frame using in-vivo magnetic resonance image (MRI) data. Emulation methods avoid computationally expensive simulations from the LV model by replacing the biomechanical model, which is defined in terms of explicit partial differential equations, with a surrogate model inferred from simulations generated before the arrival of a patient, vastly improving computational efficiency at the clinic. We compare and contrast two emulation strategies: (i) emulation of the computational model outputs and (ii) emulation of the loss between the observed patient data and the computational model outputs. These strategies are tested with two different interpolation methods, as well as two different loss functions. The best combination of methods is found by comparing the accuracy of parameter inference on simulated data for each combination. This combination, using the output emulation method (i), with local Gaussian process interpolation and the Euclidean loss function, provides accurate parameter inference in both simulated and clinical data, with a reduction in the computational cost of about 3 orders of magnitude compared to numerical integration of the differential equations using finite element discretisation techniques. 
\end{abstract}

\begin{keyword}
Left-ventricle (LV) heart model \sep Holzapfel-Ogden constitutive law \sep Magnetic Resonance Imaging (MRI) \sep Simulation \sep Emulation \sep Gaussian processes \sep Optimisation.
\end{keyword}

\end{frontmatter}

\section{Introduction}
\label{sec:Introduction}

\noindent
It is widely recognised that when 
integrated with in vivo data from cardiac magnetic resonance imaging (MRI), computational modelling of cardiac biomechanics can provide unique insights into cardiac function in both healthy and diseased states \citep{wang2015image,chabiniok2016multiphysics,gao2017changes}. For example, recent mathematical studies have demonstrated that passive myocardial stiffness is much higher in diastolic heart failure patients compared to healthy subjects \citep{xi2014understanding}. Similarly, myocardial contractility could be much higher in acute myocardial infarction (MI) patients than it is in healthy volunteers \citep{gao2017changes}. In particular, the myocardial passive properties not only affect left ventricular (LV) diastolic filling, but also influence the pumping function in heart chamber contractions (systole) through the `Frank-Starling' law \citep{Widmaier16}, the relationship between stroke volume and end diastolic volume.  

In order to comprehensively assess LV function, it is necessary to determine passive myocardial stiffness. Traditionally myocardial passive properties can be determined by a series of ex vivo or in vitro experiments \citep{dokos2002shear}. The widely used Holzapfel-Ogden (HO) constitutive law \citep{holzapfel2009constitutive} can give a detailed description of the myocardium response in passive state, including the effects of collagen fibre structure. However, determining the material parameters of this model is challenging for clinical applications, as one can not perform invasive experiments as in \cite{dokos2002shear}. One possibility of estimating these parameters non-invasively is by cardiac MRI.
The biomechanical model used in this study describes the LV dynamics during the diastolic filling process, starting from early-diastole and finishing at end-diastole, which is the point of maximum LV expansion. Both early and end-diastolic states can be measured by MRI. We can then compare, for a given patient,  these measurements to the predictions from the biomechanical model, which defines the likelihood. The biophysical parameters defining the myocardial properties (as described by the HO law) can then be inferred in an approximate  maximum likelihood sense using an iterative optimisation procedure, as discussed in  \cite{gao2015parameter}. In the context of mathematical physiology, this procedure is referred to as solving the inverse problem.

The inverse problem itself can be solved using a variety of methods and many studies have demonstrated that it is possible to estimate constitutive material parameters using in vivo measurements even with very complex constitutive relations \citep{guccione1991passive,remme2004development,sermesant2006cardiac,sun2009computationally}. However, because of the 
strong correlation among the
material parameters and sparse noisy data, the formulated inverse problem is highly non-linear \citep{xi2011myocardial,gao2015parameter}. Furthermore, determining the unknown parameters in this way is very time consuming, with the process taking days or weeks to converge, even with a modern multi-core workstation \citep{gao2015parameter,nikou2016computational}. The primary reason for this is the high computational expense of simulating from the biomechanical model, which requires a numerical integration of the underlying partial differential equations with finite element discretisation. This procedure has to be repeated hundreds or thousands of times during the iterative optimisation of the material parameters.

As a result of the high computational costs of simulating the biomechanical model, estimating myocardial properties using a process which uses this model as a simulator is not suitable for real-time clinical diagnosis. A potential approach to overcome this problem is emulation (e.g. \cite{OHagan0, Ohagan2,Ohagan1}), which has recently been explored in the closely related contexts  of cardiovascular fluid dynamics \citep{ClaytonEmu}, the pulmonary circulatory system \citep{noe-cibb} and ventricular mechanics \citep{EmuVentricularMechanics}.

Emulation methods are far more computationally efficient as most of the computation can be done in advance, making the in-clinic diagnosis faster. With emulation approaches, we simulate a large number of samples at different parameter specifications in advance and use these simulations combined with an interpolation method to replace the computationally expensive simulator in the optimisation procedure. The choice of parameter combinations from which simulations are taken can be determined effectively using a space filling design, in this case produced by a Sobol sequence \citep{Sobol67}, to spread the parameter combinations chosen in a way that aims to maximise the information about the simulator for a given number of simulations via several uniformity conditions. Optimising this design is an active research area (see e.g. \cite{DesignOpt}), which is beyond the remit of the present paper though. 

The work presented here is designed as a proof of concept study to assess the accuracy of alternative emulation strategies for learning the material properties of a healthy volunteer's LV myocardium based on only non-invasive, in vivo MRI data. To that end, we use a patient-specific model with a fixed, patient-specific LV geometry, and focus on the statistical methodology for biophysical parameter estimation. Additionally, we use a reduced parameterisation of the HO law with the biomechanical model based on the work of \cite{gao2015parameter} in MRI data. Based on this approach, we compare different emulation strategies, loss functions and interpolation methods.

The first of the emulation approaches we have tested is based on emulating the outputs of the simulator, Section~\ref{sec:EmulateOutputs}, in this case the simulated clinical data based on the described biomechanical model. Here, individual interpolators are fitted to each of the simulator outputs, using our chosen interpolation technique. We can then calculate the loss function between the predicted output of the individual models and the observed new data points from which we wish to learn the underlying myocardial properties. Minimising this loss function via a standard optimisation routine then produces estimates of the material parameters of the new subject. A variety of loss functions can be used within our emulation methods and we have compared two different ones here. The first of these is the Euclidean loss function, which assumes independence between outputs and the second is the Mahalanobis loss function \citep{Mahalanobis36} which allows for correlations.

The second emulation approach involves emulating a loss function rather than the outputs directly, Section~\ref{sec:EmulateLoss}, where again we use both the Euclidean and the Mahalanobis loss functions. For new MRI data, we calculate the loss, which quantifies the discrepancy between the model predictions and the data. Statistical interpolation is then used to obtain a surrogate loss function over the biophysical parameter space, which can be minimised with standard iterative optimisation routines.

In addition to testing these two emulation paradigms, we test two interpolation techniques based on Gaussian Processes \citep{rasmussen2006gaussian}. The first of these is a low rank Gaussian Process (GP) emulation method, which uses the complete dataset for interpolation, but uses a low rank approximation in order to scale to high dimensions \citep{Wood03}. The second method uses a local GP, where the interpolation is based on the $K$-nearest neighbours closest to the current values of the material parameters. Using a reduced number of training points from the simulations at each stage of the optimisation procedure and thereby lowering the computational costs is important, as due to the cubic computational complexity in the number of training points a standard GP would not be suitable for clinical decision support in real time.

In this work, we firstly compare different combinations of emulation methods, interpolation methods and loss functions in order to determine which method provides the best estimate of the material LV properties. We do this via a simulation study, Sections~\ref{sec:Results_interpolation}, \ref{sec:Results_emulation}, \ref{sec:Results_LossFunctions} and~\ref{sec:Results_Overall}, using additional independent simulations from the simulator as out-of-sample test data. Knowledge of the true parameter values allows us to assess the accuracy of the different combinations of methods. We then test the best combination of methods on real MRI data from the healthy volunteer from which we have taken the LV geometry, Section~\ref{sec:ResultsMRI}, to assess the accuracy of biomechanical constitutive parameter estimation in a time frame suitable for clinical applications.

\section{Left-Ventricle Biomechanical Model}
\label{sec:LVmodel}

\noindent
The LV biomechanical model describes the diastolic filling process from early-diastole to end-diastole. There are multiple different models that can be used to describe this process and these are reviewed in detail in \cite{chabiniok2016multiphysics}. The model used here is similar to those used in \cite{Wang13} and \cite{gao2015parameter}. The biomechanical model initially described in \cite{Wang13} can be thought of consisting of 5 parts: initial discretised LV geometry, the constitutive law (the HO law), the constitutive parameters, the finite element implementation, and corresponding initial and boundary conditions. Linking this biomechanical model to patient MRI data can allow the inference of unknown material parameters describing heart mechanics,\
potentially leading to improved disease diagnosis and  personalised treatments \citep{Gao17}.

The mathematical model takes 3 inputs: the initial discretised LV geometry constructed from MRI images at early-diastole (Section~\ref{sec:LVmodel_geometry}), corresponding initial and boundary conditions (Section~\ref{sec:LVmodel_BoundaryConditions}), and constitutive parameters (Section~\ref{sec:LVmodel_HOlaw}). Based on these inputs, the mathematical model, implemented in ABAQUS\footnote{ABAQUS (Simulia, Providence, RI USA)}, simulates the diastolic filling process using the HO law (Section~\ref{sec:LVmodel_HOlaw}) and a finite element implementation \citep{gao2015parameter}. The output of the mathematical model then gives a model of the LV state at end-diastole, which can be compared to the corresponding in-vivo MRIs. These MRIs at end-diastole are used to measure circumferential strains taken at 24 locations\footnote{These are based on the American Heart Association definition as in \cite{gao2015parameter}.} and the end diastolic volume. These measurements can be compared against those generated by the biomechanical model for various constitutive parameters in order to learn the parameters associated with the volunteer from which the MRI were taken.

Each simulation from the mathematical model without parallelisation takes about 18 minutes on our local Linux workstation\footnote{Intel(R) Xeon(R) CPU, 2.9GHz, 32G memory}, or around 4.5 minutes with parallelisation on 6 CPUs. Note that the 18 or 4.5 minutes are required for just a single parameter adaption step of an iterative optimisation, or a single addition to the emulator.



\subsection{Initial discretised LV geometry}
\label{sec:LVmodel_geometry}

\begin{figure}
    \centering
    \begin{subfigure}[b]{0.25\textwidth}
        \includegraphics[width=\textwidth]{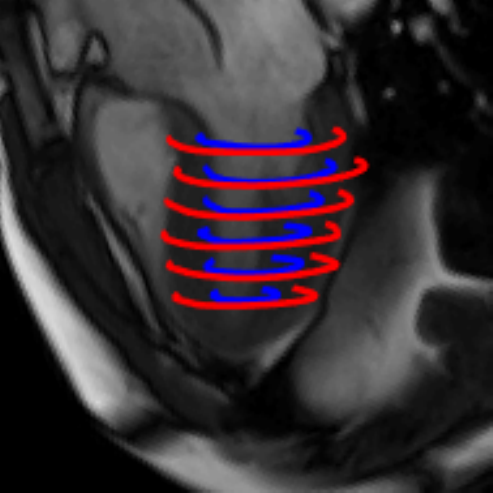}
        \caption{}
				\label{fig:LV_model_construction_a}
    \end{subfigure}
    \begin{subfigure}[b]{0.34\textwidth}
        \includegraphics[width=\textwidth]{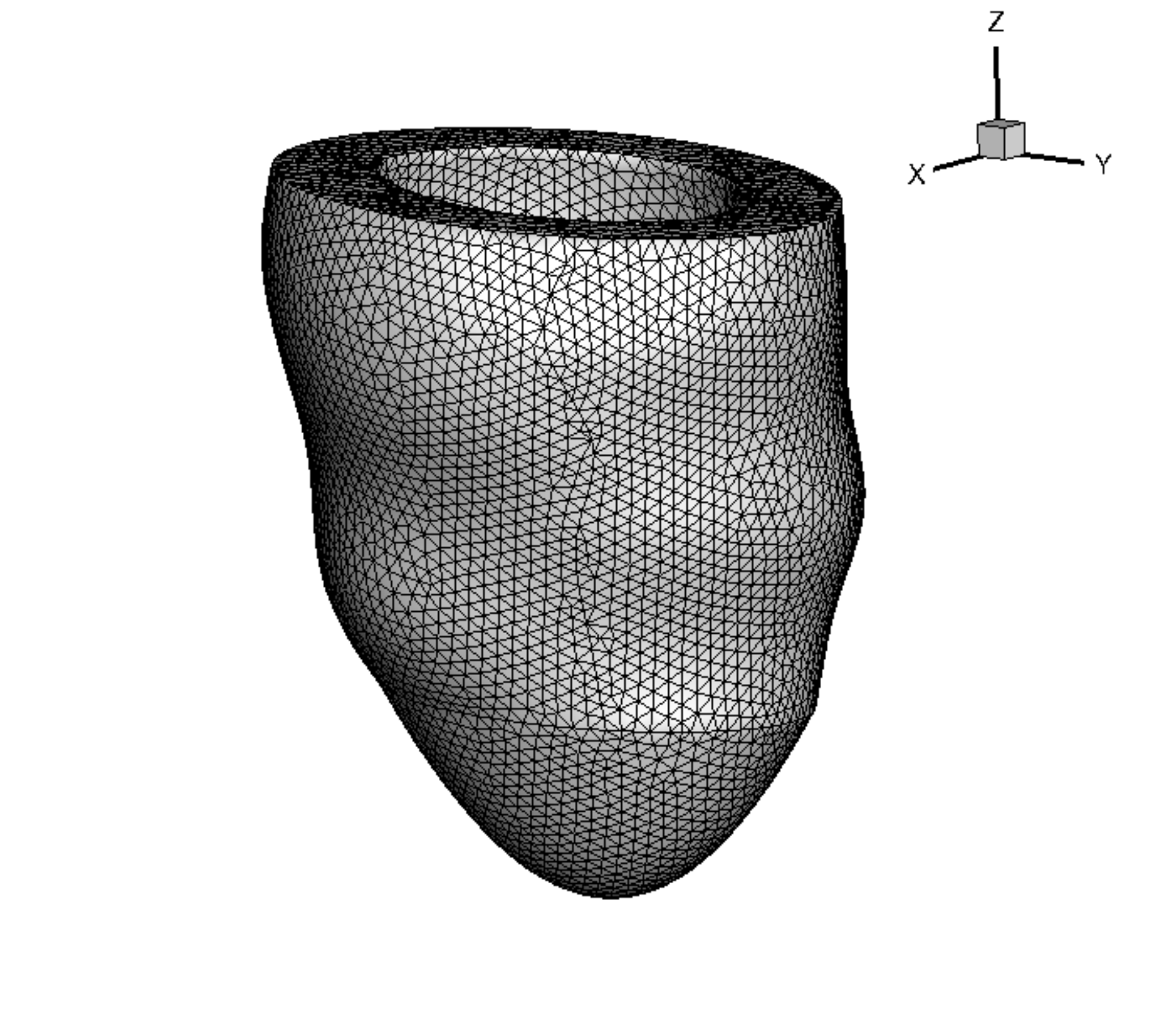}
				\caption{}
				\label{fig:LV_model_construction_b}
    \end{subfigure}
    \begin{subfigure}[b]{0.3\textwidth}
        \includegraphics[width=\textwidth]{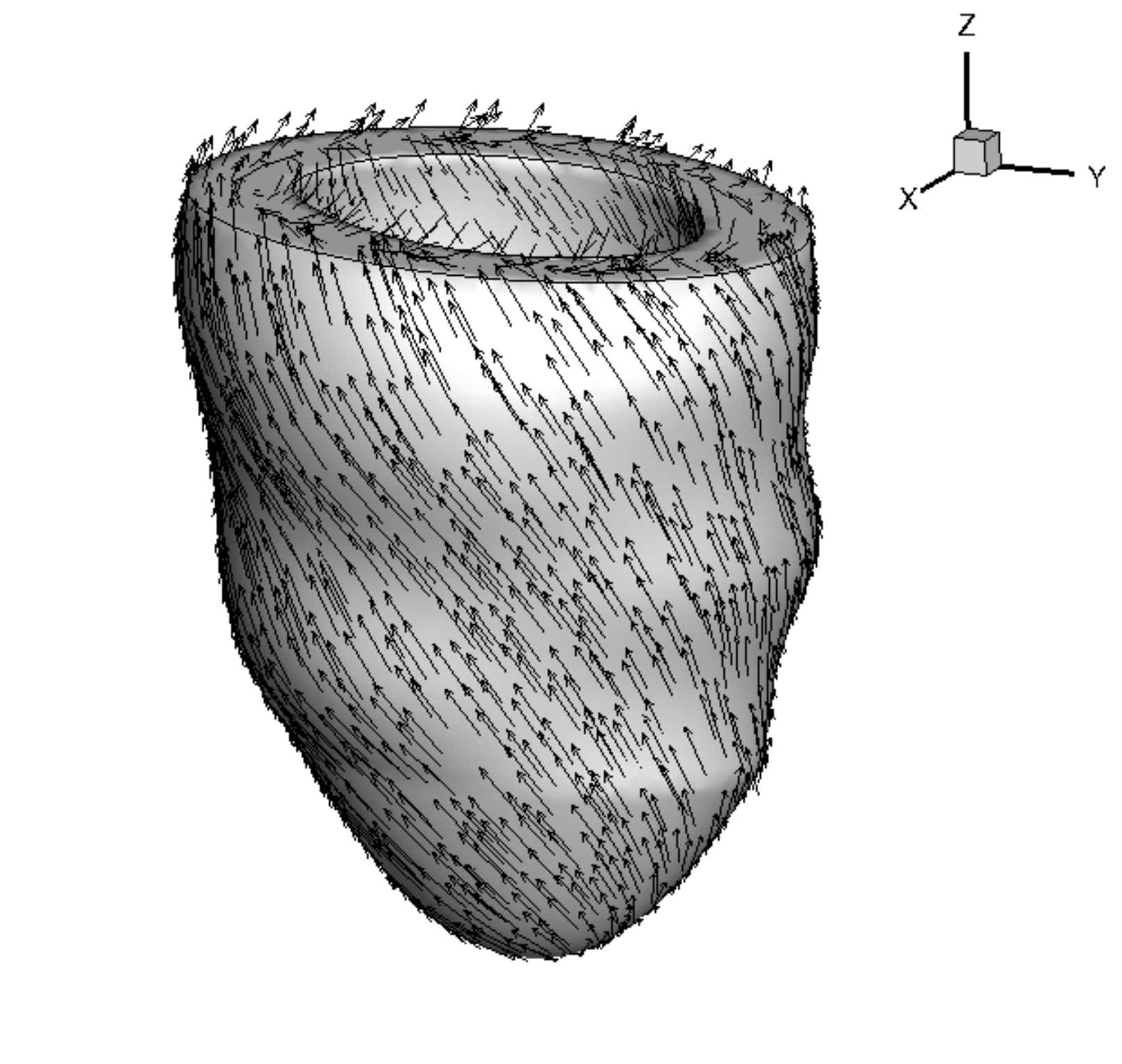}
        \vspace{-3.5mm}
				\caption{}
        \label{fig:LV_model_construction_c}
    \end{subfigure}
		\caption{The biomechanical LV model reconstructed from in vivo MRI from a healthy volunteer. (a) Segmented ventricular boundaries superimposed on a long-axis MRI; (b) the reconstructed LV geometry discretised with tetrahedron elements; (c) vector plot of fibre direction $\vec{f}$, which rotates from endocardium to epicardium.}
		\label{fig:LV_model_construction}
\end{figure}

\noindent
The initial discretised LV geometry can be obtained through constructing a 3D model based on the MRI scans \citep{Wang13}. The scans consist of a series of 6 or 7 short-axis cine images which cover the ventricle\footnote{The MRI study was conducted on a Siemens MAGNETOM Avanto (Erlangen, Germany) 1.5-Tesla scanner with a 12-element phased array cardiac surface coil.  Cine MRI images were acquired using the steady-state precession imaging protocol. Patient consent was obtained before the scan.}.
For each cardiac cycle there are usually around 35 frames from end-diastole to early-diastole. The images of the early-diastole are then used to create the initial discretised LV geometry, while the end-diastole images will provide the final measurements of the circumferential strains and the LV volume. To create the discretised LV model, the endocardial (inner) and epicardial (outer) boundaries of the LV are segmented from cine images at early-diastole as done in \cite{Gao14}, e.g. Figure~\ref{fig:LV_model_construction_a}. A 3D model of the LV can then be constructed in Solidworks\footnote{Solidworks (Dassault Systems SolidWorks Corp., Waltham, MA USA)}, e.g. Figure~\ref{fig:LV_model_construction_b}. Finally, Figure~\ref{fig:LV_model_construction_c} is constructed using a rule-based fibre-generation method, see \cite{gao2014dynamic}, giving us the initial discretised LV geometry used in the biomechanical model. In the context of the present study, we consider this a fixed input and focus our work on developing parameter inference methods rather than a tool that can work for all possible subjects. Extensions to allow for different LV geometries is the subject of future work.

\subsection{Initial and Boundary Conditions}
\label{sec:LVmodel_BoundaryConditions}

\noindent
The initial and boundary conditions, in particular LV pressure, play an important role in myocardial dynamics.
Unfortunately, blood pressure within the cavity of the left ventricle can only be measured invasively, by direct catheter measurement within the LV cavity.
Due to potential complications and side effects, these measurements are not available for healthy volunteers. We have therefore fixed the boundary conditions, including the pressure, at values considered sensible for healthy subjects,  based on the work of \cite{bouchard1971evaluation}.

\subsection{Constitutive Law}
\label{sec:LVmodel_HOlaw}

\noindent
The final part of the biomechical model is the constitutive law for characterising the material properties of myocardium. In this study, we use the invariant-based constitutive law \citep{holzapfel2009constitutive}, based on the following strain energy function:
\begin{equation}
\begin{split}
		\Psi &= \frac{a}{2b} \{ \text{exp}[b(I_1-3)]-1\} +\sum_{i \in \{f,s\}} \frac{a_i}{2b_i}\{ \text{exp}[b_i(I_{4i}-1)^2]-1\} \\
		& + \frac{a_\text{fs}}{2b_\text{fs}}[\text{exp}(b_\text{fs}I_\text{8fs}^2)-1] + \half K (J-1)^2,
		\label{HO}
\end{split}
\end{equation}
in which $a$, $b$, $a_\text{f}$, $b_\text{f}$, $a_\text{s}$, $b_\text{s}$, $a_\text{fs}$ and $b_\text{fs}$ are unknown material parameters, $I_1, I_{4i}$, and $I_\text{8fs}$ are the invariants corresponding to the matrix and fibre structure of the myocardium, which are  calculated as 
$$
I_1 = \mbox{trace}(\CC),\quad
I_\text{4f} = \vec{f}_0 \cdot (\CC \vec{f}_0) ,\quad
I_\text{4s} = \text{s}_0 \cdot (\CC \text{s}_0) ,\quad
I_\text{8fs} = \fhg_0 \cdot (\CC \text{s}_0) 
$$
in which $\fhg_0$ and $\text{s}_0$ are the myofibre and sheet orientations, which are determined through a rule-based approach \citep{Wang13} and are known before the simulation (initial conditions). 
$\CC$ is the right Cauchy-Green deformation tensor, defined as $\CC=\FF^T\FF$, where 
$\FF$ is the deformation gradient describing the motion of myocardium
and hence how its shape changes in 3D with time. The term $\half K (J-1)^2$ accounts for the incompressibility of the material, where $K$ is a constant ($10^6$) and $J$ is the determinant of $\FF$. The HO law forms a major part of the biomechanical model, and the 8 constitutive parameters, $a$, $b$, $a_\text{f}$, $b_\text{f}$, $a_\text{s}$, $b_\text{s}$, $a_\text{fs}$ and $b_\text{fs}$, are unknown inputs into the model, which we wish to learn. The  
accuracy of parameter estimation for real data can be based on stretch-stress curves, as discussed in Section~1 of the online supplementary materials.
	
However, it has previously been found in \cite{gao2015parameter} that the 8 parameters are strongly correlated, which suggests that a model reduction is advisable to ensure identifiability. The authors further demonstrated that myofibre stiffness, the parameter most relevant for clinical applications, can be estimated from in vivo data with a reduced parameterisation; see Section~2 of the online supplementary materials. In fact, \cite{hadjicharalambous2016non} even estimated passive myocardial stiffness using a reduced form of the HO law with only a single unknown parameter. In the present study, similarly to \cite{gao2015parameter}, we group the eight parameters of \eqref{HO} into four, so that: 
\begin{equation}
\begin{aligned}
a&=\theta_1\, a_0,  &b&=\theta_1 \, b_0 \\ 
a_\text{f}&=\theta_2 \,a_\text{f0}, &a_\text{s}&=\theta_2 \, a_\text{s0} \\ 
b_\text{f}&=\theta_3 \, b_\text{f0}, &b_\text{s}&=\theta_3 \, b_\text{s0} \\ 
a_\text{fs}&=\theta_4 \, a_\text{fs0},  &b_\text{fs}&=\theta_4 \, b_\text{fs0}
\label{eq::theta}
\end{aligned} 
\end{equation}
where $\theta_i \in [0.1,5]: i=1,2,3,4$  are the parameters to be inferred from in vivo data, and $a_0$, $b_0$, $a_\text{f0}$, $a_\text{s0}$, $b_\text{f0}$, $b_\text{s0}$, $a_\text{fs0}$, and $b_\text{fs0}$ are reference values from the published literature \citep{gao2017changes}\footnote{The reference values are, up to 2 decimal places: $a_0=0.22$, $b_0=1.62$, $a_\text{f0}=2.43$, $a_\text{s0}=0.56$, $b_\text{f0}=1.83$, $b_\text{s0}=0.77$, $a_\text{fs0}=0.39$, and $b_\text{fs0}=1.70$.}. 
Our results obtained with this dimension reduction are consistent with the experimental results reported in \cite{dokos2002shear}.



\section{Statistical Methodology}
\label{sec:Methods}

\noindent
This section reviews the notion of a simulator and emulator, as well as establishing the notation that is used throughout the rest of the paper. It also provides details about the different emulation strategies that are going to be used in this paper, as well as the different interpolation methods considered.

\subsection{Simulation}
\label{sec:Simulators}

\noindent
A \emph{simulator}, $ \bs{\ms{m}} $, is a mathematical model that relies on a computationally expensive numerical solution of the underlying systems equations.
In the present study, the mathematical model is the soft-tissue mechanical description of the left ventricle based on the Holzapfel-Ogden strain energy function, as discussed in the previous section. The numerical procedure is the finite element discretization of the resulting partial differential equations.
 The inferential process, i.e. estimating the unknown inputs or parameters $\bftheta_0$ underlying the observed clinical data $ \y_0 $, is computationally expensive and infeasible in settings where solutions are required within a short time frame, for instance in the context of clinical decision support. The prohibitive computational time that makes inference challenging is due to the time needed for a single (\emph{forward}) \emph{simulation} from the computational model, where by forward simulation we mean generating a (possibly multivariate) output $\y = (y_1, \ldots, y_J) = \bs{\ms{m}}(\bftheta) $ for a given 
parameter vector or
input 
$ \bftheta $.
In the context of the present study, $J=25$, and the outputs $y_i$ are the 24 circumferential strains and the LV volume at end of diastole, as predicted by the mathematical model.

Given our clinical data, $\y_0$, 
which are the measured circumferential strains and the end-of-diastole LV volume obtained from MRI,
we can estimate the unknown parameter vector $\bftheta_0$ by finding the corresponding input to the simulator which gives rise to an output which is as close as possible to the observed clinical data, $ \y_0 $. While our clinical data is assumed to come from the same data generating process $ \bs{\ms{m}} $ for an unknown input, $\bftheta_0 $, in practise there will be a systematic deviation due to noisy measurement and model mismatch. The simplest approach to estimating the unknown input or parameter vector $\bftheta$  is to choose the loss function as the negative log-likelihood:
\begin{equation}\label{eq:trueloss}
\ell(\bftheta| \bs{\ms{m}},\y_0) = \alpha d( \bs{\ms{m}} (\bftheta), \y_0 ) + Z,
\end{equation}
for a given metric function $ d $ measuring the distance between a simulation $ \y =\bs{\ms{m}}(\bftheta) $ and data $ \y_0 $, and some positive constants $\alpha$ and $Z$.  We can then estimate the input to the model by minimising the true loss in \eqref{eq:trueloss}:
\begin{equation}
\label{key}
\hat{\bftheta} = \arg \min_{\bftheta} \ell(\bftheta| \bs{\ms{m}},\y_0),
\end{equation}
effectively giving us the maximum likelihood estimate of $\bftheta$. This method becomes prohibitive if a single simulation exceeds a certain amount of time, as it does with the biomechanical model considered in the present work. The numerical procedure based on finite element discretization
requires approximately 18 minutes for a single simulation, or 4.5 minutes with parallelisation on 6 CPUs on our computing system\footnote{Dual Intel Xeon CPU E5-2699 v3, 2.30GHz, 36 cores and 128GB memory.} Any optimisation of the true loss, \eqref{eq:trueloss}, would require the evaluation of the simulator at every iteration of the optimisation routine, potentially hundreds or thousands of times, with each iteration taking between 4.5 and 18 minutes. This is computationally limiting if we wish to use the method for clinical decision support in real time.

\subsection{Emulation}
\label{sec:Emulation}

\noindent
An \emph{emulator} is a statistical model that is a cheap and fast approximation to the true computational model (simulator), $ \bs{\ms{m}} $, in this case the biomechanical model. It is used to replace the simulator in order to speed up both computations and inference, and it is also referred to as a \emph{meta-model} (or \emph{surrogate model}) as it represents a model of a model. An emulator can be built using any interpolation technique such as regression splines, polynomial regression, Gaussian processes, etc; see Section~\ref{sec:interpolation} for more details. Once a method has been chosen and the emulator has been fitted to the training  data, we will denote it as $ \hat{\bs{\ms{m}}} $.

In order to fit a statistical model and replace the simulator, we need training data from the simulator itself in the form of simulations $ \D = \{ (\bftheta_1, \y_1), \dots,$ $ (\bftheta_N, \y_N) \} = \{ \bs{\Theta}, \bs{Y} \}$. 
In the context of the present application, the input vectors $\bftheta_i$ are the biomechanical parameter vectors discussed in Section~\ref{sec:LVmodel_HOlaw}.
These inputs into the simulator, $\bs{\Theta}$, are chosen based on a space filling design, using Sobol sequences.
These so-called low-discrepancy sequences are known to lead to improved convergence in the context of quasi-Monte Carlo; see e.g. \cite{QMC}. A more efficient coverage of the input space is possible using more advanced statistical design methods, as e.g. discussed in \cite{DesignOpt}, but these explorations are beyond the remit of present work.

The outputs of the simulator, $\bs{Y}$, are the resulting clinical values based on the assumed data generating process, $\bs{\ms{m}}$. 
In the present application, the output vectors $\y_i$ are the vectors of 24 circumferential strains and LV volume at end of diastole.
Whilst generating large numbers of simulations is computationally expensive, this can be massively parallelised in advance and before the arrival of the patient at the clinic.

Previously, given the clinical data, $\y_0$, and a simulator, $\bs{\ms{m}}$, we could not estimate the unknown input, $\bftheta_0$, using the loss function (negative log-likelihood) given in~\eqref{eq:trueloss} fast enough for effective use within a clinical environment. This was due to the high simulation time required for each single input. Now, however, we can replace the true loss function in~\eqref{eq:trueloss} with a surrogate loss function, $\ell$, based on an emulation method; see Section~\ref{sec:emulation_methods} for details. Minimisation of the surrogate loss (surrogate negative log-likelihood) for any metric function $ d $ will be fast and suitable for real-time precision medicine, as it does not involve any simulation from the computationally expensive model.

We can use a variety of different metric functions within our surrogate loss, $\ell$. The most obvious of these is the Euclidean norm, 
$\| \hat{\bs{\ms{m}}}(\bftheta) - \y_0 \|^2$. Under the assumption of independent and identically (iid) normally distributed errors (i.e. deviations of the clinical data from the emulator outputs) with zero mean and variance $\sigma^2$, the Euclidean loss function is equivalent to the negative log-likelihood, up to a scaling factor and an additive constant $Z(\sigma)$:
\begin{equation} \label{eq:surrogatelosseuclidean}
\ell(\bftheta| \hat{\bs{\ms{m}}},\y_0) = \frac{1}{2\sigma^2} \| \hat{\bs{\ms{m}}}(\bftheta) - \y_0 \|^2 + Z(\sigma).
\end{equation}
An extension of the Euclidean loss which allows for a correlations between the outputs is the Mahalanobis loss function:
\begin{equation} \label{eq:surrogatelossmahalanobis}
\ell(\bftheta| 
\hat{\bs{\ms{m}}},\y_0) =
\frac{1}{2}(\hat{\bs{\ms{m}}}(\bftheta) - \y_0 )^\top \bs{\Sigma}^{-1} (\hat{\bs{\ms{m}}}(\bftheta) - \y_0 ) + Z(\bs{\Sigma}),
\end{equation}
which is equivalent to the negative log-likelihood of a multivariate Gaussian distribution with covariance matrix $\bs{\Sigma}$ up to a constant, $Z(\bs{\Sigma})$.
To minimise the computational costs at the clinic, the covariance matrix is pre-computed from the training data,
$\bs{\Sigma}=\mr{cov}(\bs{Y})$,
and then kept fixed. Its main purpose is to allow for the spatial correlations between the 24 circumferential strains at different locations on the LV.

\subsection{Emulation Frameworks}
\label{sec:emulation_methods}

\subsubsection{Output Emulation}
\label{sec:EmulateOutputs}

\begin{algorithm}[t]
\caption{Inference using an emulator of the outputs}
\label{alg:outputemulation}
	\begin{algorithmic}[1]
		\State Simulate from the model $ \bs{\ms{m}}(\bftheta_1), \dots, \bs{\ms{m}}(\bftheta_N) $ at space filling inputs $ \bftheta_1, \dots, \bftheta_N $. 
		\State Fit $ J $ independent real-valued emulators $ \hat{\bs{\ms{m}}} = (\hat{\m}_1, \ldots, \hat{\m}_J)  $, one for each of the $ j = 1, \dots, J $ outputs of the simulator.
		\State Given data $ \y_0 $ and the emulator, $\hat{\bs{\ms{m}}}$, construct the surrogate-based loss function $ \ell(\bftheta \mid \hat{\bs{\m}}, \y_0) $
		\State Minimize the surrogate-based loss function to give the estimates, $\hat{\boldsymbol\theta}_0$.
\end{algorithmic}
\end{algorithm}

\noindent
Emulating the outputs of the simulator, the LV model, involves fitting multiple individual models, one for each of the $J$ outputs of the simulator, $\bs{\ms{m}}$. These outputs, $y_j: j =1, \ldots,J$, are fitted using the inputs of the simulator, $\bs{\Theta}$, with an appropriate interpolation method; see Sections~\ref{subsec:heartLocalGP} and~\ref{subsec:heartLowRankGP}. Given the multiple independent models, $ \hat{\bs{\ms{m}}} = (\hat{\m}_1, \ldots, \hat{\m}_J)  $, estimates of the parameter vector, $\hat{\bftheta}_0$, can be found for any new set of outputs $\y_0$ by minimising the difference between $\y_0$ and $\hat{\bs{\ms{m}}}(\boldsymbol\theta)$ with a loss function:
\begin{equation}
		\label{eq:Outputs_Loss}
		\hat{\bftheta}_0 = \arg \min_{\boldsymbol\theta} \ell(\bftheta \mid \hat{\bs{\m}}, \y_0).
\end{equation}
The loss function, $\ell$, in \eqref{eq:Outputs_Loss} can take a variety of forms, including the Euclidean and the Mahalanobis loss functions given in \eqref{eq:surrogatelosseuclidean} and \eqref{eq:surrogatelossmahalanobis}. An algorithmic description of the output emulation method is given in Algorithm~\ref{alg:outputemulation}.

The advantage of emulating the outputs is that the statistical models can be fitted in advance, before the data have been collected from the clinic, meaning that when a patient comes into the clinic, an estimation of the biomechanical parameter vector $\hat{\bftheta}_0$ can be carried out relatively quickly. The disadvantage is that multiple potentially correlated model outputs must be fitted,
leading to higher computational costs at training time than emulating the loss function directly.


\subsubsection{Loss Emulation}
\label{sec:EmulateLoss}

\noindent
An alternative strategy 
is \emph{loss emulation}. This entails direct emulation of the losses 
$ \ell_n = \ell(\bftheta_n| \bs{\ms{m}} ,\y_0)$ 
rather than the simulator outputs $ \y_n = \bs{\ms{m}}(\bftheta_n) $, for $ n = 1, \dots, N $. Given simulations $ \D = \{ \bs{\Theta}, \bs{Y} \} $ and clinical data $ \y_0 $, it is possible to evaluate the loss function at each of the training points and data $\y_0 $ and record the obtained score. To follow this approach we fit a single real-valued emulator to training data:
\begin{equation} \label{eq:loss1}
	\mathcal{D}_{\ell} = \{ (\bftheta_n, \ell_n) \colon n = 1, \dots, N \},
\end{equation}
where 
$ \ell_n = \ell(\bftheta_n| \bs{\ms{m}} ,\y_0)$ 
is the loss function, for a given metric $d$, evaluated at the $n$th design point from the corresponding simulation output,
$ \y_n = \bs{\ms{m}}(\bftheta_n) $. The metric $ d $ should be chosen according to the problem, and it can capture the correlation between the model outputs. Now it is possible to fit a single real-value emulator $ \hat{\ell}(\bftheta| \bs{\ms{m}},\y_0) $ of $ \ell(\bftheta| \bs{\ms{m}},\y_0) $ based on the training data, $ \mathcal{D}_\ell $, using a single statistical model instead of 
a vector of model outputs.
Estimation of the parameters can now be done cheaply by minimizing the emulated loss function:
\begin{equation}
\label{eq:loss2}
	\hat{\bftheta}_0 = \arg \min_{\boldsymbol\theta}  \mathbb{E} \{ \hat{\ell}(\bftheta| \bs{\ms{m}},\y_0) \}.
\end{equation}
where $\mathbb{E} $ denotes the conditional expectation predicted by the interpolation method, in our case the conditional mean of a Gaussian process.
An algorithmic description of the loss emulation method is given in Algorithm~\ref{alg:lossemulation}. For further illustration, an additional example,  on the Lotka-Volterra system, can be found in \cite{noephd}.

The advantage of loss emulation over output emulation is a reduction of the training complexity, as a multi-dimensional vector 
is replaced by a scalar as the target function.
The disadvantage is that, as opposed to output emulation, the emulator can only be trained \emph{after} the 
patient has come into the clinic and the training data have become available. This implies that on production of the training data, 
the emulator has to be trained and the resulting emulated loss function has to be optimized, leading to higher computational costs at the time a clinical decision has to be made.
However, these computational costs are still low compared to running the simulator.

Loss emulation is closely related to Bayesian optimization, reviewed e.g. in \cite{BayesOpt} and \cite{noephd}, which is a strategy to iteratively include further query points by 
 trading off exploration versus exploitation via some heuristic or information-theoretic criterion.
However, every additional query point requires a computationally expensive simulation from the mathematical model, which prevents fast clinical decision making in real time
and renders Bayesian optimization infeasible for the purposes of our study.

\subsection{Interpolation Methods}
\label{sec:interpolation}

\noindent
We have considered several interpolation methods, based on Gaussian processes (GPs).
GPs have been widely used in the context of emulation; see e.g. \cite{OHagan0, Ohagan2,Ohagan1}. For a comprehensive introduction to GPs, the reader is referred to \cite{rasmussen2006gaussian}.
Each of the interpolation methods can be used with both of the emulation paradigms described in the previous section, Section~\ref{sec:emulation_methods}.

\begin{algorithm}[t]
\caption{Inference using an emulator of the losses}
\label{alg:lossemulation}
	\begin{algorithmic}[1]
		\State Simulate from the model $ \bs{\ms{m}}(\bftheta_1), \dots, \bs{\ms{m}}(\bftheta_N) $ at space filling inputs $ \bftheta_1, \dots, \bftheta_N $. 
        \State Calculate the set of loss functions $ \ell(\bftheta_n \mid \bs{\ms{m}},  \y_0) $, for $ n = 1, \dots, N $, between each individual simulation and the observed data $ \y_0 $.
		\State Emulate the losses using a single real-valued model $ \hat{\ell}(\bftheta \mid \bs{\ms{m}}, \y_0) $
		\State Estimate $ \hat{\bftheta}_0 $ by minimizing the mean of the loss-emulator $ \mathbb{E}\{ \hat{\ell}(\bftheta \mid \bs{\ms{m}}, \y_0) \} $
\end{algorithmic}
\end{algorithm}

\subsubsection{Local Gaussian Process\label{subsec:heartLocalGP}}

\noindent
When the sample size $N$ is large, it is not feasible to use exact
GP regression on the full dataset,
due to the $O(N^{3})$ computational complexity of the $N\times N$
training covariance matrix $\boldsymbol{K}$ inversion. A possible
approach is to use sparse GPs as in \citet{Titsias2009}, which considers
a fixed number of $m$ inducing variables $\boldsymbol{u}=(u_{1},\dots,u_{m})$,
with $m\ll N$, corresponding to inputs $\boldsymbol{Z}=[\boldsymbol{z}_{1},\dots,\boldsymbol{z}_{m}]^{\T}$.
The locations of the inducing points and the kernel hyperparameters
are chosen with variational inference, i.e. 
by maximizing  a lower bound on the log marginal likelihood, which can be derived by applying Jensen's inequality.
The computational costs of this approach are $O(Nm^{2})$. Initially we
tried sparse GPs with 100, 500 and 1000 inducing points but, using
the code accompanying the paper by \citet{Titsias2009}, the prediction
time was between 0.5 and 0.6 seconds for 100 inducing points, around
one second for 500, and in the order of a few seconds for 1000 inducing
points\footnote{Dual Intel Xeon CPU E5-2699 v3, 2.30GHz, 36 cores and 128GB memory.}. This means that minimization of the surrogate-based loss
would still be slow as approximately 1 second is required for a single
evaluation. The optimization time would exceed two and a half hours
for 500 inducing points when using $10,000$ function evaluations. With the cost of variational sparse GP models with larger numbers of inducing points being so large, we  can only use about 100 inducing points in order to keep to our goal of real-time in-clinic decision making. However, using such few inducing points was found to lead to around a quarter of the outputs of the biomechanical model being poorly predicted.

With the performance of the variational sparse GPs being poor when the number of inducing points are selected to give a clinically relevant decision time, we instead use a local GP approach based on the $K$-nearest-neighbours instead \citep{Gramacy2015}. This method uses the standard GP prediction formulas described in \cite{rasmussen2006gaussian}, but subsetting the training data. Whenever we require a prediction
at a given input, we find the training inputs representing the $K$-nearest-neighbours
in input-domain, which will form the local set of training inputs,
and the corresponding outputs will represent the local training outputs.
Note that every time we ask for a prediction at a different input,
the training sets need to be re-computed and the GP needs to be trained
again. However, because of the small number of neighbours $K \ll 1000$
usually selected, this method is computationally fast and accurate;
see \citet{Gramacy2015} for a discussion.

\citet{Gramacy2015} further discuss adding a fixed number of distant
points in order to help in the estimation of the length scale parameters,
but this comes with extra computational costs required by the iterative
choice of which point to add to the set of neighbours. Given the time
limitations required by our goal (real-time clinical decision support
systems) we do not pursue this approach. Furthermore, this is mostly
relevant when the interest lies in building predictive models able
to make good predictions when the training data are distant from each
other. Since we are working on a compact set which is densely
covered by the Sobol sequence, this is less relvant. For generic
training data $\D=\{(\boldsymbol{\theta}_{1},y_{1}),\dots,(\boldsymbol{\theta}_{N},y_{N})\}=\{\boldsymbol{\Theta},\boldsymbol{y}\}$,
we give an algorithmic description in Algorithm~\ref{alg:gp_prediction}.

In Algorithm~\ref{alg:gp_prediction}, the $K\times K$ training covariance matrix is
$\boldsymbol{K}=[k(\boldsymbol{\theta}_{i}',\boldsymbol{\theta}_{j}')]_{i,j=1}^{K}$, the $K\times1$ vector of covariances between the training points
and the test point is $\boldsymbol{k}(\boldsymbol{\theta}_{*})=(k(\boldsymbol{\theta}_{1}',\boldsymbol{\theta}_{*}),\dots,$ $k(\boldsymbol{\theta}_{K}',\boldsymbol{\theta}_{*}))$
and $\boldsymbol{m}=(m(\boldsymbol{\theta}_{1}'),\dots,m(\boldsymbol{\theta}_{K}'))$
is the $K\times1$ prior mean vector. We consider a constant mean
function $m(\boldsymbol{\theta})=c$.
For the kernel $k(.,.)$ we choose the Automatic Relevance Determination Squared Exponential kernel (see e.g. \cite{rasmussen2006gaussian}),
as widely used in the emulation of computer codes literature; see e.g. \citet{Fang2006,Santner2003}.
The kernel hyperparameters are the output scale (determining the function variance) and the input length scales, one length scale for each dimension.
These hyperparameters are
estimated by maximizing the log marginal
likelihood using the Quasi-Newton method. The standard deviation of the additive Gaussian noise,
$\sigma$, is initialized at a small value, $\sigma=10^{-2}$, to reflect the fact that
the mathematical model of the LV is deterministic\footnote{Even for deterministic models, a small non-zero value for $\sigma$ is usually assumed, to avoid numerical instabilities of the covariance matrix inversion.}

\begin{algorithm}[t]
\caption{Predicting from a local Gaussian process at $\boldsymbol{\theta}_{*}$}
\label{alg:gp_prediction}
	\begin{algorithmic}[1]
\State Find the indices $\mc N(\boldsymbol{\theta}_{*})$ of the points in $\boldsymbol{\Theta}$
having the $K$ smallest Euclidean distances from $\boldsymbol{\theta}_{*}$;
\State Training inputs: $\boldsymbol{\Theta}_{K}(\boldsymbol{\theta}_{*})=\{\boldsymbol{\theta}_{1}',\dots,\boldsymbol{\theta}_{K}'\}=\{\boldsymbol{\theta}_{i}:i\in\mc N(\boldsymbol{\theta}_{*})\}$;
\State Training outputs: $\boldsymbol{y}_{K}(\boldsymbol{\theta}_{*})=\{y_{1}',\dots,y_{K}'\}=\{y_{i}:i\in\mc N(\boldsymbol{\theta}_{*})\}$;
\State Train a GP using the data $\D_{K}(\boldsymbol{\theta}_{*})=\{\boldsymbol{\Theta}_{K}(\boldsymbol{\theta}_{*}),\boldsymbol{y}_{K}(\boldsymbol{\theta}_{*})\}$;
\State Predictive mean: $\hat{f}(\boldsymbol{\theta}_{*})=m(\boldsymbol{\theta}_{*})+\boldsymbol{k}(\boldsymbol{\theta}_{*})^{\T}[\boldsymbol{K}+\sigma^{2}\boldsymbol{I}]^{-1}(\boldsymbol{y}_{K}(\boldsymbol{\theta}_{*})-\boldsymbol{m})$;
\State Predictive variance: $s^{2}(\boldsymbol{\theta}_{*})=k(\boldsymbol{\theta}_{*},\boldsymbol{\theta}_{*})-\boldsymbol{k}(\boldsymbol{\theta}_{*})^{\T}[\boldsymbol{K}+\sigma^{2}\boldsymbol{I}]^{-1}\boldsymbol{k}(\boldsymbol{\theta}_{*})$.
\end{algorithmic}
\end{algorithm}

The CPU time required to get a prediction from the local Gaussian
process is approximately 0.18 seconds\footnote{Dual Intel Xeon CPU E5-2699 v3, 2.30GHz, 36 cores and 128GB memory.}
using the $K=100$ nearest neighbours of a given point. The number
of neighbours $K$ needs to be selected on the basis of the computational
time allowed to reach a decision in a viable time frame, but keeping
in mind that $K$ also controls the accuracy of the emulation. In
our experiments we found that $K=100$ was sufficiently fast for the method to be applicable in the clinic while leading to accurate predictions at the test inputs, as discussed below in the Results section.

For this method, the surrogate-based loss and the emulated loss were optimized
using the Global Search algorithm by \citet{Ugray2007}, implemented
in MATLAB's Global Optimization toolbox.\footnote{Available from \url{https://uk.mathworks.com/products/global-optimization.html}.
We use the default choice of 2000 trial points and 400 stage one points. Consider running
a local solver from a given starting point $\bftheta_{0}$,
ending up at the point of local minimum $\hat{\bftheta}$. The
\emph{basin of attraction} corresponding to that minimum is defined
as the sphere centred at $\hat{\bftheta}$ and having radius
equal to $\|\bftheta_{0}-\hat{\bftheta}\|$. All starting
points falling inside the sphere are assumed to lead to the same local
minimum $\hat{\bftheta}$, hence no local solver is run and
they are discarded. In simple words, stage one of the Global Search
algorithm scatters initial points in the domain and scores them from
best to worst by evaluating the function value and constraints. Then
an interior-point local solver \citep{Byrd2000} is run from each
trial point, starting from the one that was scored best (lowest function
value), and excluding points that fall into the basins of attraction
of previously found minima. When all the stage one points have been
analyzed, stage two generates more random points and the same procedure
is run a second time.}

\subsubsection{Low-Rank Gaussian Processes\label{subsec:heartLowRankGP}}

\noindent
Along with local GPs based on the $K$-nearest-neighbours, described
in \prettyref{subsec:heartLocalGP}, we report results for another
type of statistical approximation: low-rank GPs, as described in Section
5.8.2 of \citet{wood17}, whose main ideas are summarized here for
generic training data $\D=\{(\boldsymbol{\theta}_{1},y_{1}),\dots,(\boldsymbol{\theta}_{n},y_{n})\}=\{\boldsymbol{\Theta},\boldsymbol{y}\}$.

Let $\boldsymbol{C}=\boldsymbol{K}+\sigma^{2}\boldsymbol{I}$ be the
$n\times n$ covariance matrix of $\boldsymbol{y}$ and consider its
eigen-decomposition $\boldsymbol{C}=\boldsymbol{U}\boldsymbol{D}\boldsymbol{U}^{\T}$
with eigenvalues $|D_{i,i}|\geq|D_{i+1,i+1}|$. Denote by $\boldsymbol{U}_{k}$
the submatrix consisting of the first $k$ eigenvectors of $\boldsymbol{U}$,
corresponding to the top $k$ eigenvalues in $\boldsymbol{D}$. Similarly,
$\boldsymbol{D}_{k}$ is the diagonal matrix containing all eigenvalues
greater than or equal to $D_{k,k}$. \citet{wood17} considers replacing
$\boldsymbol{C}$ with the rank $k$ approximation $\boldsymbol{U}_{k}\boldsymbol{D}_{k}\boldsymbol{U}_{k}^{\T}$
obtained from the eigen-decomposition. Now, the main issue is how
to find $\boldsymbol{U}_{k}$ and $\boldsymbol{D}_{k}$ efficiently
enough. A full eigen-decomposition of $\boldsymbol{C}$ requires $O(N^{3})$
operations, which somewhat limits the applicability of the rank-reduction
approach. A solution is to use the Lanczos iteration method to find
$\boldsymbol{U}_{k}$ and $\boldsymbol{D}_{k}$ at the substantially
lower cost of $O(N^{2}k)$ operations, see Section B.11 in \citet{wood17}.
Briefly, the algorithm is an adaptation of power methods to obtain
the truncated rank $k$ eigen-decomposition of an $N\times N$ symmetric
matrix in $O(N^{2}k)$ operations. However, for large $N$, even $O(N^{2}k)$
becomes prohibitive. In this scenario the training data are randomly
subsampled by keeping $n_{r}$ inputs and an eigen-decomposition is
obtained for this random selection with $O(n_{r}^{2}k)$ computational
cost. 

We used the implementation found in the \emph{R} package
\texttt{mgcv} by \citet{wood17}, with the following settings: $n_{r}=2000$
(the package default), $k=2000$ for output emulation, while $k=1000$
for loss emulation. The kernel used was an isotropic Mat\'{e}rn 3/2 kernel,
with lengthscale set to the default of \citet{KammannWand03}: $\lambda=\max_{ij}\|\boldsymbol{\theta}_{i}-\boldsymbol{\theta}_{j}\|$.
The remaining model hyperparameters are estimated by maximizing the
log marginal likelihood. The final model used an interaction term between each of the 4 model parameters, as well as a second interactive term between the inverses of the model parameters:
\begin{align}
		\label{eq:GAM_inverse}
		\tilde{\y}_j \sim \beta_j\textbf{1} + f(\boldsymbol\theta) + f(\boldsymbol{\tau}) + \boldsymbol\varepsilon ~~\mbox{for}~~ j = 1,\dots,J
\end{align}
where $\boldsymbol{\tau} = 1/\boldsymbol\theta$, $f(\boldsymbol\theta) \sim \ms{GP}_{\ms{LR}}(m(\boldsymbol\theta),K(\boldsymbol\theta,\boldsymbol\theta'))$, $f(\boldsymbol{\tau}) \sim \ms{GP}_{\ms{LR}}(m(\boldsymbol{\tau}),K(\boldsymbol{\tau},\boldsymbol{\tau}'))$ and $\ms{GP}_{\ms{LR}}(\cdot)$ denotes a low rank GP.
The model specification with the two interaction terms was found to reduce the variation in the predictive accuracy as the volume increases and the strains decrease. This can be seen in the predictions of the test and training data in Figure~2 and~3 of the online supplementary materials.

Minimization of the surrogate-based loss $\ell(\cdot \mid \hat{\bs{\ms{m}}}, \y_0)$ and
the emulated loss $\hat{\ell}(\cdot \mid \bs{\ms{m}}, \y_0)$ is performed by the Conjugate
Gradient method implemented in the \emph{R} function \texttt{optim} \citep{Nash90},
with maximum number of iterations set to 100. To avoid being trapped
in local minima, 50 different starting points from a Sobol sequence
were used. The best minimum found was kept as the estimate, discarding
the remaining 49 optima.

\subsubsection{Multivariate-output Gaussian Processes}
\label{sec:multivariate_emulation}

\noindent
The previous two subsections have focussed on single-output GPs, while potentially correcting for the correlation structure of the outputs via a modified objective function, using the Mahalanobis distance  defined in \prettyref{eq:surrogatelossmahalanobis}.
One can model the correlation structure between the outputs directly via
\begin{equation}
\mathbf{Cov}[\outVec(\inVec_i),\outVec(\inVec_j)] \; = \; \Cij(\inVec_i,\inVec_j)\COmatrix
\end{equation}
where $K(\inVec_i,\inVec_j)$ is the covariance between $y_k(\inVec_i)$ and $y_k(\inVec_j)$ for any output $k$, and $\COmatrix$ is a matrix of the covariances between the outputs, i.e. the circumferential strains and the LV volume.
Various approaches have been proposed in the literature. The approach taken in \cite{Ohagan1} and \cite{Ohagan2} is to place a non-informative prior on $\COmatrix$ and integrate $\COmatrix$ out in the likelihood. This leads to a closed-form solution in terms of a matrix-normal distribution; see \cite{Ohagan1} and \cite{Ohagan2} for explicit expressions. 
However, we found that in combination with Algorithm~\ref{alg:gp_prediction} -- to deal with the $O(\Ndata^3)$ computational complexity -- the computational costs of running the emulator were in the order of hours, rather than minutes, which renders this approach not viable for clinical decision support in real time. 

An alternative approach is to explicitly model the correlation structure of the outputs via
\begin{equation}
\mathrm{Cov}[\outSca_k(\inVec_i),\outSca_l(\inVec_j)] \; = \; \Cij(\inVec_i,\inVec_j)\COij(\bfu_k,\bfu_l)
\end{equation}
taking into account covariates $\bfu_k$ and $\bfu_l$ associated with the $k$th and $l$th outputs, $y_k$ and $y_l$, respectively. \cite{SteveRobertsGP} pursue this approach in the context of time series analysis, where $\bfu_k$ and $\bfu_l$ are scalar variables indicating different time points. In our application, $\bfu_k$ and $\bfu_l$ are vectors indicating the locations on the surface of the left ventricle associated with the circumferential strains. Due to the highly non-Euclidean geometry of this space, the choice of kernel is not obvious. A naive approach that we tried is to project the locations onto a linear space defined by the first principal component \citep{YunlongHuangMSc}. The results were not encouraging,  due to the information loss incurred by the map. Future work could try projections onto nonlinear maps, like Hilbert curves \citep{HilbertCurve1,HilbertCurve2}, generative topographic maps \citep{GTM}, or self-organising maps \citep{SOM}. 

A further alternative is the method of \cite{AlvarezLawrence09,LawrenceMultiOutGP}, who have proposed sparse convolved GPs for multi-output regression. Their method assumes that there is an underlying process which governs all of the outcomes of the model and treats it as a latent process. Modelling this latent process as a GP leads to a GP prior over the outputs, inducing cross covariance between the outputs and effectively introducing correlations between them. We can use the interpolation method of \cite{AlvarezLawrence09,LawrenceMultiOutGP} within either of the emulation frameworks introduced in Section~\ref{sec:emulation_methods}. There are however problems with doing this: training a convolved GP with $N$ training points requires the inversion of a $DN \times DN$ matrix (where $D=25$ is the number of outputs) which is currently infeasible with all of the training data ($N=10,000$), even when choosing the number of inducing points using the method proposed in \cite{AlvarezLawrence09,LawrenceMultiOutGP}. Instead we can choose a strategy similar to that proposed in Section~\ref{subsec:heartLocalGP}. This again, however, proves to be computationally expensive as fitting a single local emulator requires more than 15 minutes\footnote{Intel Xeon CPU E5-606,2.13GHz}, without consideration of the computational costs of the subsequent optmization of the LV model parameters. When this is included within either of the emulation methods (Algorithms~\ref{alg:outputemulation} and~\ref{alg:lossemulation}), the time becomes too large for a clinical decision support system, as it is infeasible to make a prediction within a clinically relevant time frame. 

Since the focus of our study is to develop an emulation framework for a clinical decision support system that can work in real time, we have restricted our analysis to the univariate methods described in Sections~\ref{subsec:heartLocalGP} and~\ref{subsec:heartLowRankGP}.

\section{Data and Simulations}
\label{sec:Data}

\noindent
For training the emulator, we used 10,000 parameter vectors generated from a Sobol sequence \citep{Sobol67} in a compact 4-dimensional parameter space, with
$\theta_1, \ldots, \theta_4 \in [0.1,5]^4$, where the parameter bounds reflect prior knowledge available from \cite{gao2015parameter}. The 4-dimensional parameter vectors 
are then transformed to the original 8-dimensional parameter space using the transformation (\ref{eq::theta}).
The 8-dimensional parameter vectors are then inserted into the HO strain energy function (\ref{HO}). Following the finite element discretisation method described in 
\cite{Wang13}, the soft-tissue mechanical equations are numerically solved to produce a 25-dimensional output vector associated with each parameter vector; these are 24 circumferential strains and the LV volume at end of diastole. The Sobol sequence is extended to generated an independent test set of additional 100 parameter vectors, for which the same procedure is followed to associate them with output vectors of circumferential strains and LV volume. As a real data set, we used 24 circumferential strains and the LV volume at end of diastole obtained from the cardiac MRI images of a healthy volunteer, following the procedure described in \cite{gao2015parameter}. 

\section{Results}
\label{sec:Results}

\noindent
To summarise, we have introduced two emulation frameworks which can be used to infer the parameters of the LV biomechanical model; see Sections~\ref{sec:EmulateOutputs} and~\ref{sec:EmulateLoss}. We have applied these methods with two different loss functions, the Mahalanobis loss function and the Euclidean loss function, and two different interpolation methods, low rank GPs and local GPs; see Sections~\ref{subsec:heartLocalGP} and~\ref{subsec:heartLowRankGP}. Testing each combination of these methods means that there is a total of 8 different alternative procedures.

\begin{figure}[t]
				\centering
        \begin{subfigure}[b]{0.47\textwidth}
                \centerline{\includegraphics[width=\columnwidth]{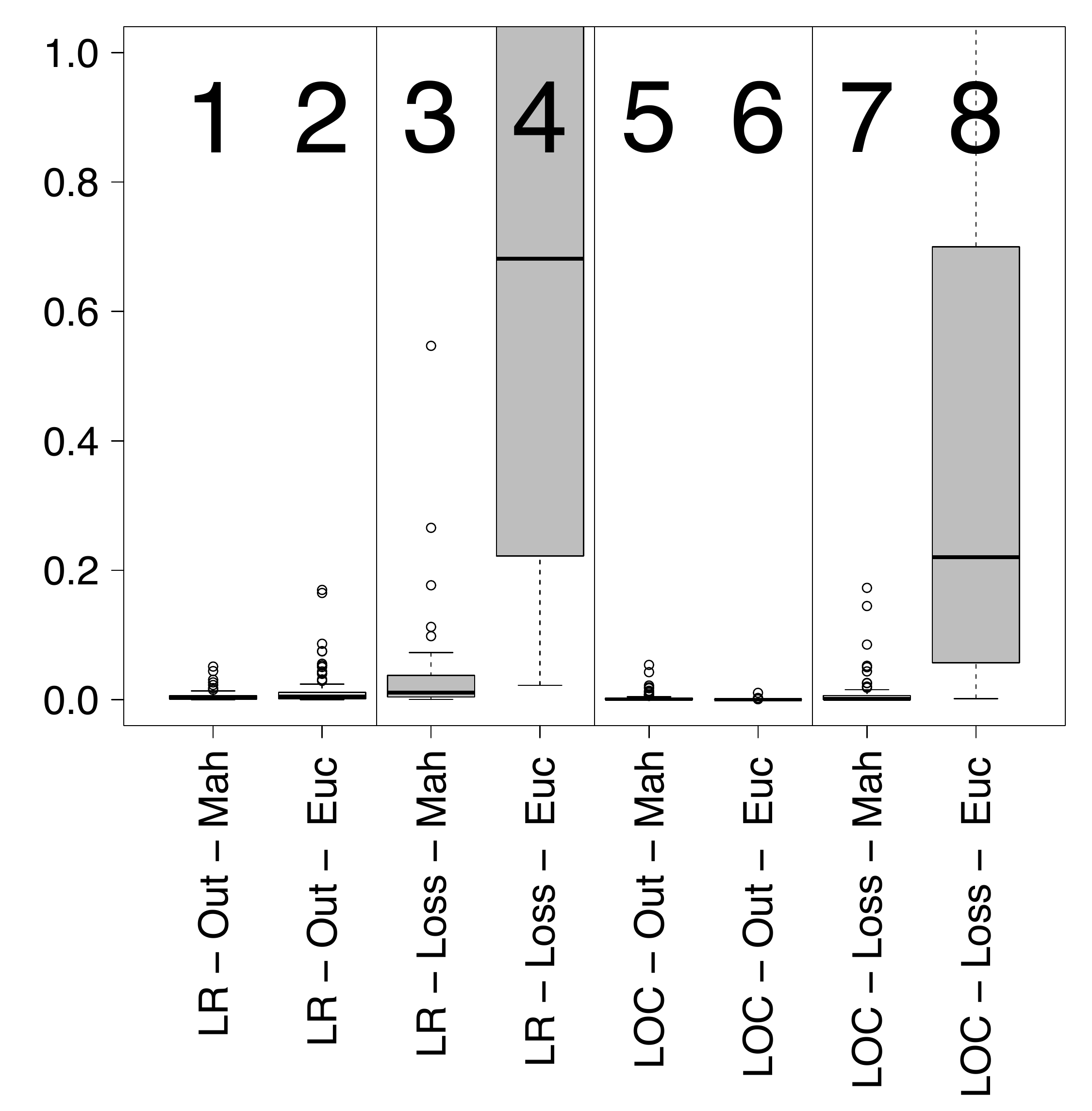}}
                \caption{ }
                \label{fig:Boxplot_Allparam_Sq_ZO}
        \end{subfigure}
        \begin{subfigure}[b]{0.47\textwidth}
                \centerline{\includegraphics[width=\columnwidth]{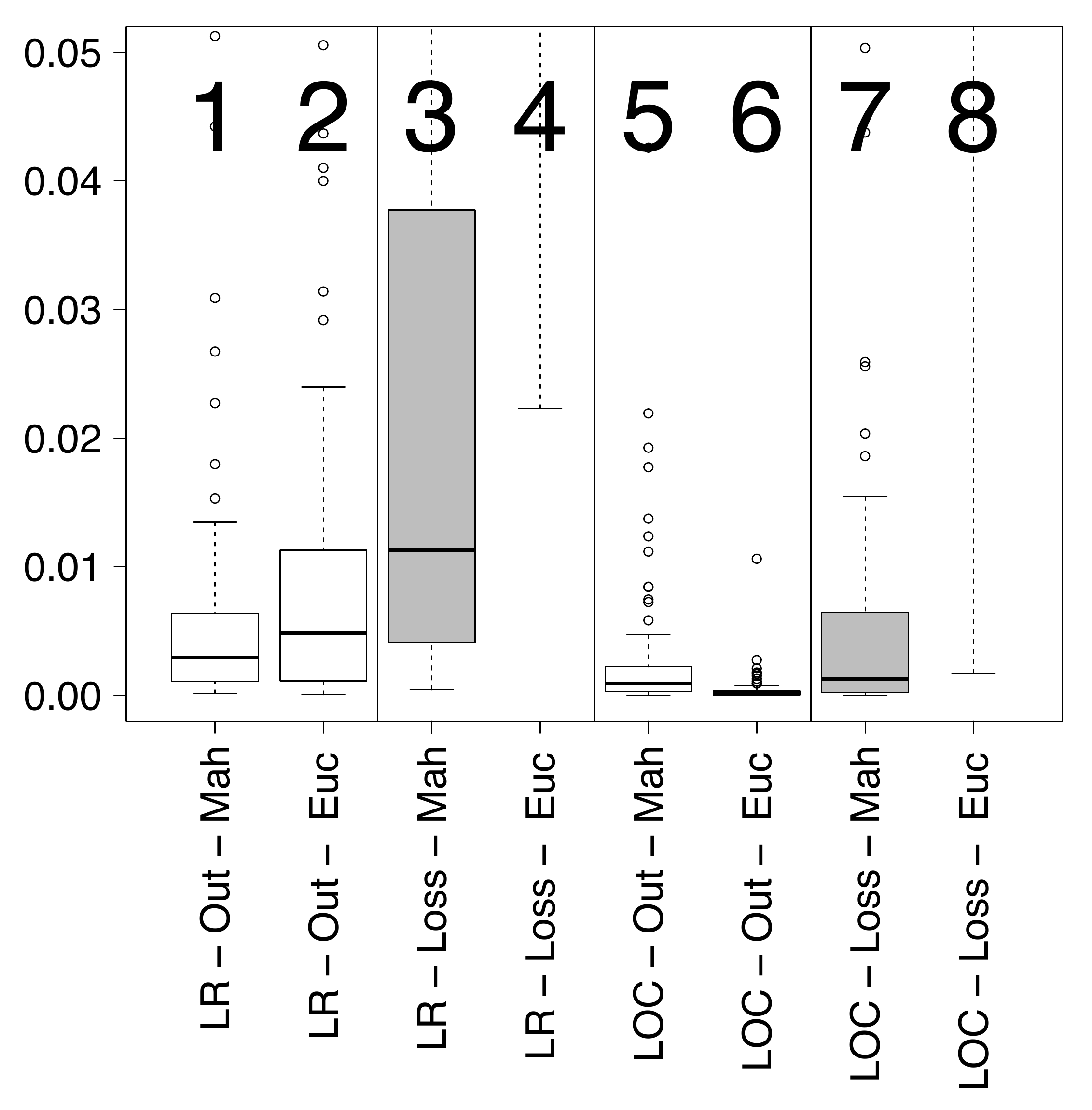}}
                \caption{ }
                \label{fig:Boxplot_Allparam_Sq_ZI}
        \end{subfigure}
        \caption{Boxplots of the mean squared error distribution in the prediction of all the model parameters. Panel (a)~shows boxplots of the mean squared error in parameter-space for all the 8 methods, panel (b) shows the same boxplots but with a reduced scale on the y-axis. The methods from left to right on each plot are as follows: low rank GP (LR) output emulation (Out) with Mahalanobis loss function (Mah) and Euclidean loss function (Euc), LR-GP loss emulation (Loss) with Mahalanobis loss function and Euclidean loss function, local GP (LOC) output emulation with Mahalanobis loss function and Euclidean loss function, and LOC loss emulation with Mahalanobis loss function and Euclidean loss function.
        The outliers are due to non-convergence of the optimization algorithm and the strong correlation between the parameters of the HO law.
        }
				\label{fig:Results_Simulations_Boxplot_SquaredError}
\end{figure}

We have applied and assessed the proposed methods in a two-pronged approach. Firstly, in Sections~\ref{sec:Results_interpolation}, \ref{sec:Results_emulation}, \ref{sec:Results_LossFunctions} and~\ref{sec:Results_Overall}, we have tested the 8 different combinations of methods on synthetic data, where the true parameter values of the underlying biomechanical model are known;
see the previous section for details on how the training and test data were generated.
We compare the methods using the Mean Square Error (MSE).\footnote{Note that the likelihood is computationally expensive and intractable. Hence, we do not compare the methods using the log posterior of the parameters, as this would involve approximations, using e.g. variational methods or expectation propagation, and this would have to be repeated 100 times (the number of test data) at high computational costs.} The distribution of 100 MSEs is given in Figure~\ref{fig:Results_Simulations_Boxplot_SquaredError} and summarised with the median and the (1st, 3rd) quartiles in Table~\ref{tab:Results}, representing 3 of Tukey's five number summary.\footnote{We do not present plus or minus the interquartile range as this can lead to the wrong impression of a negative MSE.}

Finally, we have applied the method with the best performance in Section~\ref{sec:ResultsMRI} to clinical data generated from a healthy volunteer's cardiac MRI scan, where we can compare our performance against the gold standard results of \cite{gao2015parameter}. 

\subsection{Comparison of Interpolation Methods}
\label{sec:Results_interpolation}

\noindent
Looking at the two interpolation methods, the local GP method (boxplots 5-8 in  Figure~\ref{fig:Results_Simulations_Boxplot_SquaredError}) outperforms the low rank GP method (boxplots 1-4 in Figure~\ref{fig:Results_Simulations_Boxplot_SquaredError}). The reason for the difference in performance between the two methods is the size of the noise variance that is estimated. With the low rank GP method, a larger noise variance is estimated as the interpolation must fit to the entire dataset. The larger variance of the errors is in mismatch with the deterministic nature of process that we are modelling. Instead of estimating the variance with maximum likelihood, one could consider a Bayesian approach that discourages larger values of the variance with a restrictive prior. 
However, besides the confounding effect of switching statistical inference paradigms (from maximum likelihood to Bayesian inference), the available code for this is not available in the \texttt{mgcv} package in \emph{R}.

Conversely, with the local GP method, a much smaller error variance is estimated, which more closely matches the deterministic data generation method. This is a result of there only being a small number of points that the interpolant must fit. These points are local, giving more detail of the local surface than the low rank GP method, which uses a selected number of points from across the whole dataset.

\begin{table}[t]
		\caption{Table giving the median (1st quartile, 3rd quartile) of the mean squared error (in parameter-space) in the prediction of all the model parameters. 
		The considered interpolation methods (Interp. Meth.) are Low-Rank GPs and Local GPs, the target of the emulation (Emulation Target) is either the model output or the loss, and two loss functions are compared: Euclidean and Mahalanobis.
		The method with the best predictive performance, the output emulation method with local GP interpolation and the Euclidean loss function, is given in bold.}
		\centering
    \begin{tabular}{ c  c  c  c }
    Interp. Meth. 	& Emulation Target  & Euclidean & Mahalanobis \\ \hline
		Low-Rank GP & Output 		& 0.0048 (0.0012,0.0107) &  0.0030 (0.0011,0.0062) \\
		Low-Rank GP & Loss 			& 0.6814 (0.2222,1.5234) &  0.0113 (0.0041,0.0377) \\ \hline
		Local GP 		& Output 		& \textbf{0.0001 (0.0000,0.0003)} &  0.0009 (0.0003,0.0022) \\
		Local GP 		& Loss 			& 0.2201 (0.0588,0.6777) &  0.0013 (0.0002,0.0063)
		\end{tabular}
		\label{tab:Results}
\end{table}

\subsection{Comparison of Emulation Frameworks}
\label{sec:Results_emulation}

\noindent
Out of the two emulation frameworks, the output emulation method (boxplots 1, 2, 5 and 6 in Figure~\ref{fig:Results_Simulations_Boxplot_SquaredError}) gives the most accurate parameter estimates, outperforming the loss emulation method (boxplots 3, 4, 7 and 8 in Figure~\ref{fig:Results_Simulations_Boxplot_SquaredError}) for all interpolation methods and loss functions. The output emulation method provides accurate estimates for all the different combinations of interpolation methods and loss functions, while the loss emulation method provides poor estimates in some cases. The improved parameter estimation of the output emulation method is a result of using multiple separate emulators. These multiple emulators better model the complex non-linear relationships between the parameters and the outputs than it is possible with the single emulator used with the loss emulation method. In the loss emulation method, the differences between the patient data and the simulations are summarised in one loss function, reducing the performance of the method.

\subsection{Comparison of Loss Functions}
\label{sec:Results_LossFunctions}

\noindent
In terms of the accuracy of the parameter inference, the Euclidean loss and Mahalanobis loss perform differently in different emulation methods. Firstly, for the loss emulation method the Mahalanobis loss function (boxplots 3 and 7 in Figure~\ref{fig:Results_Simulations_Boxplot_SquaredError}) clearly outperforms the Euclidean loss function (boxplots 4 and 8 in Figure~\ref{fig:Results_Simulations_Boxplot_SquaredError}) in all cases. The reason for the difference is that the loss function summarises how similar the patient data is to the simulations and this is done more realistically by the Mahalanobis loss function in this case.  This is because there are  spacial correlations between the outputs due to measuring the circumferential strains at different neighbouring locations on the left ventricle. The Mahalanobis loss function accounts for this through including a correlation estimate, whereas the Euclidean loss function does not.

In comparison to the loss emulation method, for the output emulation method it is less clear which loss function gives the best results. The Mahalanobis loss function is marginally better 
for the low rank GP method (boxplot 1 is better than boxplot 2 in Figure~\ref{fig:Results_Simulations_Boxplot_SquaredError}), 
while the Euclidean loss function gives the best performance for the local GP method (boxplot 6 is better than boxplot 5 in Figure~\ref{fig:Results_Simulations_Boxplot_SquaredError}). The reason for the Euclidean loss function performing best for the local GP method is because of potential inaccuracies in the correlation matrix used for the Mahalanobis loss function. Firstly, the correlation matrix is a global measure based on the whole dataset and may not accurately represent the true correlations between the local points due to limited numerical precision\footnote{Using a local correlation matrix was also tested, but limited accuracy and numerical stability of the correlation matrix due to using only a small number of local points meant that the performance did not improve over the global correlation matrix.}. 
Secondly, this is aggravated by lack of numerical stability when inverting the covariance matrix.
Thirdly, the loss function minimised in the output emulation method is based on the errors between the emulators and patient data, whereas the correlation matrix has been calculated based on only the patient data. 

\subsection{Overall Best Method in Simulation Study}
\label{sec:Results_Overall}

\noindent
In conclusion, the results of our simulation study show the following. (1) The local GP method outperforms the low rank GP method and is the better of the two interpolation methods. (2) The best emulation method is the output emulation method and this outperforms the loss emulation method in all the different combinations of interpolation method and loss function tested. (3) The Mahalanobis loss function gives the best performance for the loss emulation method. (4) For the output emulation method, the Mahalanobis method is marginally better 
for the low rank GP method, but for the local GP method the Euclidean loss function gives the best parameter estimates. (5) Overall, the simulation study results show that the best performing combination of methods is the output emulation method, using the local GP as the interpolation method and the Euclidean loss function (boxplot 6 in Figure~\ref{fig:Results_Simulations_Boxplot_SquaredError}). This combination of methods will be used on the cardiac MRI data of the healthy volunteer in Section~\ref{sec:ResultsMRI}.

\begin{figure}[t]
				\centering
        \begin{subfigure}[b]{0.47\textwidth}
                \centerline{\includegraphics[width=\columnwidth]{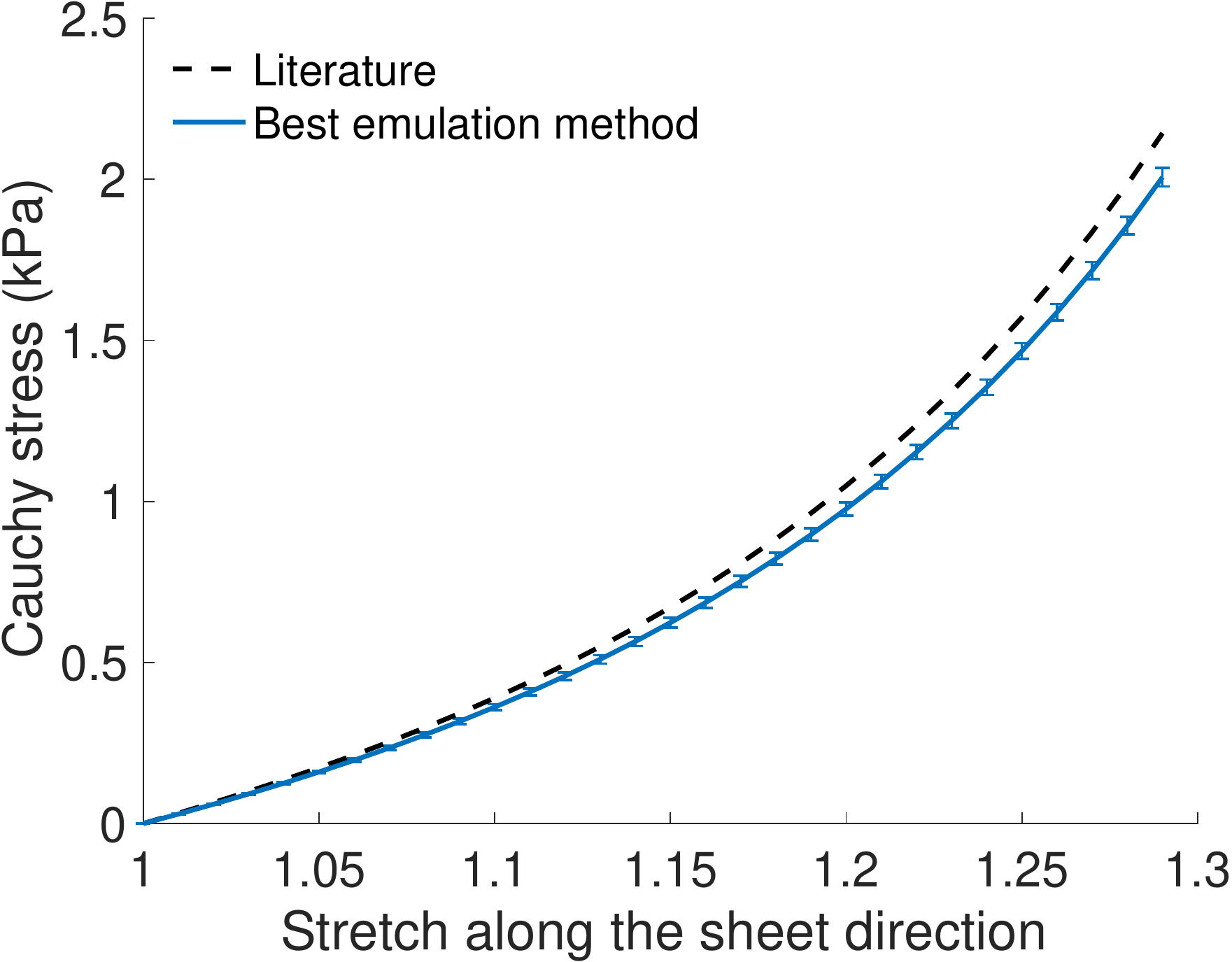}}
                \caption{ }
                \label{fig:RealResults_a}
        \end{subfigure}
        \begin{subfigure}[b]{0.47\textwidth}
                \centerline{\includegraphics[width=\columnwidth]{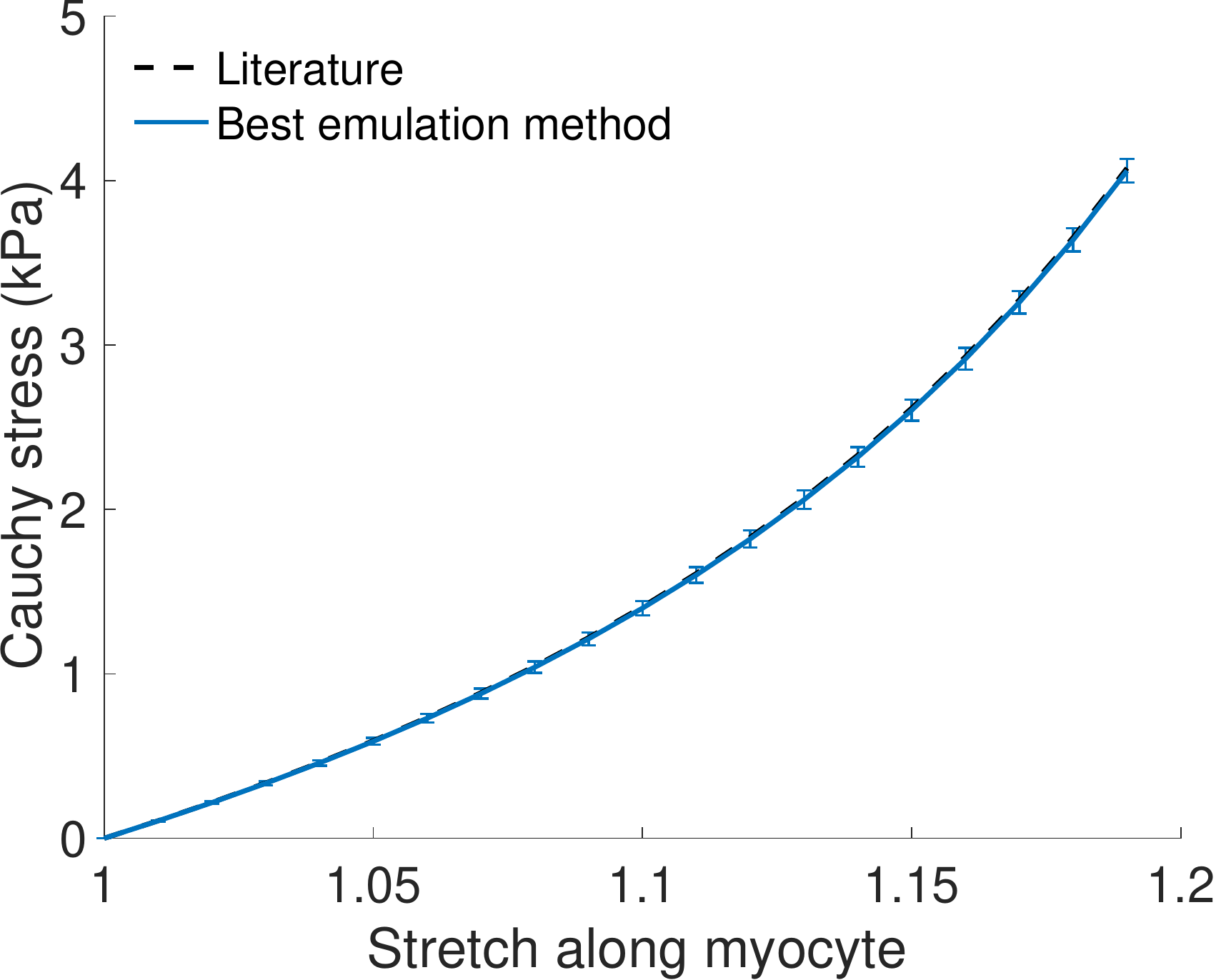}}
                \caption{ }                
                \label{fig:RealResults_b}
        \end{subfigure}
        \caption{Plots of the Cauchy stress against the strech along (a) the sheet direction and (b) the myocyte direction. Literature curves are taken from the gold standard method in \cite{gao2017changes} and are given as dashed black lines. Estimates of the curves from the best emulation method, the emulation of the outputs method combined with the local GP interpolation method and the Euclidean loss function, are given as a blue solid line. 
        The 95\% confidence intervals, shown as error bars,
        are approximated using the sampling method described in Section~\ref{sec:ResultsMRI}.}
				\label{fig:RealResults}
\end{figure}

\subsection{Application to Cardiac MRI Data}
\label{sec:ResultsMRI}

\noindent
Figure~\ref{fig:Results_Simulations_Boxplot_SquaredError} and Table~\ref{tab:Results} show that the method which gives the most accurate parameter prediction is the emulation of the outputs method combined with the local GP interpolation and the Euclidean loss function. We have applied this strategy to estimate the material parameters for the heart model of a healthy volunteer described in Section~\ref{sec:LVmodel}, using the set of 24 circumferential strains and the LV cavity volume extracted from cardiac MRI images, as described in Section~\ref{sec:Data}.
The true model parameters are not known in this case, so as opposed to the simulation study we do not have a proper `gold standard' for evaluation.  We therefore use the following alternative procedure.
We first estimate the constitutive parameters with the method of \cite{gao2015parameter,gao2017changes}, that is, with the method using the computationally expensive simulator. From these parameters, we calculate the stretch-stress relationships along the directions of the sheets and the myocytes, following the procedure described in  \cite{holzapfel2009constitutive}.
We use these graphs as a surrogate `gold standard', which we compare with the corresponding graphs obtained from the parameters obtained with our emulation approach.

Figure~\ref{fig:RealResults} shows, as dashed lines, the estimate of the stretch-stress relationship for the healthy volunteer using the `gold standard' method of \cite{gao2015parameter,gao2017changes}. 
For comparison, the solid blue lines show 
the estimates of the stress-stretch relationship obtained from the best emulation method identified in the previous sections,
Sections~\ref{sec:Results_interpolation}--\ref{sec:Results_Overall},
the emulation of the outputs method combined with the local GP interpolation method and the Euclidean loss function. 

For uncertainty quantification, we numerically estimated the Hessian at the minimum surrogate loss (\ref{eq:surrogatelosseuclidean}). Its inverse represents an approximate lower bound on the variance-covariance matrix in parameter space\footnote{The Hessian is the empirical Fisher information matrix. The lower bound would be exact (Cramer-Rao lower bound) if we could take an expectation with respect to the data distribution. Recall that saying that matrix {\bf A} is a lower bound on matrix {\bf B} means that {\bf B} - {\bf A} is positive semi-definite.}. The uncertainty in the estimate can then be obtained by sampling from a multivariate normal distribution, with the covariance set to the inverse of the Hessian, $\mathrm{MVN}(\hat{\bftheta},\textbf{H}(\hat{\bftheta})^{-1})$, and calculating the corresponding confidence  intervals.

The results in Figure~\ref{fig:RealResults} show that the emulation method accurately estimates the stretch-stress relationship in the myocyte direction. The agreement between the `gold standard' and the prediction with our emulation method is nearly perfect, with a deviation that is less than the predicted single-standard deviation width. For the stretch-stress relationship in the sheet direction, the agreement is also very good, although the deviation exceeds the predicted standard deviation in this case. 
A possible explanation is that parameter sensitivity in the sheet directions is very low when only using regional circumferential strains and the LV cavity volume to formulate the objective function, as reported in \cite{gao2015parameter}, thus the uncertainty of estimating the stiffness in the sheet direction will be higher than that in the myocyte direction. It is expected that higher accuracy will be achieved when radial (transmural) strains are included when inferring the parameters.
While the differences between the stretch-stress curves obtained with the simulator and our emulator are minor, there is a substantial difference in the computational costs. 
For the simulator, that is, the original procedure described in \cite{gao2015parameter,gao2017changes}, the computational costs are in the order of over a week.
The estimation procedure with the proposed emulator, on the other hand, could be carried out in less than 15 minutes\footnote{Dual Intel Xeon CPU E5-2699 v3, 2.30GHz, 36 cores and 128GB memory.}, giving us a reduction of the computational complexity by about three orders of magnitude.


Hence, while the former procedure is only of interest in a pure research context, the latter procedure gives us estimation times that are acceptable in a clinical decision context. This is an important first step towards
bringing mathematical modelling into the clinic and making a real impact in health care.

\section{Discussion}
\label{sec:Discussion}

\noindent
We have developed an emulation framework that can be used to infer the material properties of the LV of a healthy patient in a clinically viable time frame. We have focused on developing an emulation framework that can be used in future more generalised work and have therefore tested 2 emulation methods, 2 interpolation method and 2 loss functions; see Section~\ref{sec:Methods}. Each combination of these methods has then been evaluated in a simulation study in order to determine the best method. The best method was found to be the output emulation method, using the local GP as the interpolation method and the Euclidean loss function; see Table~\ref{tab:Results}.

We have then applied the proposed emulation method to cardiac MRI data and demonstrated that it is able to accurately estimate the stretch-stress relationship  along the myocyte and sheet directions of the LV from a healthy volunteer. Our method provides a notable improvement in computational time with a speed-up of approximately 3 orders of magnitude. 
In particular, while conventional parameter estimation based on numerical simulations from the  mathematical LV model, following e.g. the approach of \cite{gao2015parameter}, leads to computational costs in the order of weeks, the proposed emulation method reduces the computational complexity to the order of a quarter hour, while effectively maintaining the same level of accuracy. This is an important step towards a clinical decision support system that can assist a clinical practitioner in real time.

A limitation of the current approach is the fact that the LV geometry is fixed. This LV geometry varies from patient to patient, and these variations need to be taken into consideration for applications to wider patient populations. We discuss how to potentially address this challenge in the next section.

\section{Future Work}
\label{sec:FurtherWork}
 
 \noindent     
The next step for this work is to design a method that is capable of fast parameter inference for multiple patients on whom we have not directly trained the emulator. For each new patient we would need to replace the single geometry used here as an input, with the new patient's data on arrival at the clinic. With no time limits on the inference, we could simply replicate this study with a different input geometry. However, in order to treat patients in a clinically viable time frame we must be able to train the emulator for the unobserved patient before they enter the clinic. We can do this by using simulations from multiple different LV geometries as our training data. Low-dimensional representations of each geometry can then be included as variables in the interpolation method of the emulator and we can learn how these changes affect the output of the biomechanical model. When new patient data then arrives, these low dimensional representations can be calculated and included in the loss function, which must be minimised in the emulation method.

\begin{figure}[t]
				\centering
        \begin{subfigure}[b]{0.32\textwidth}
                \centerline{\includegraphics[width=\columnwidth]{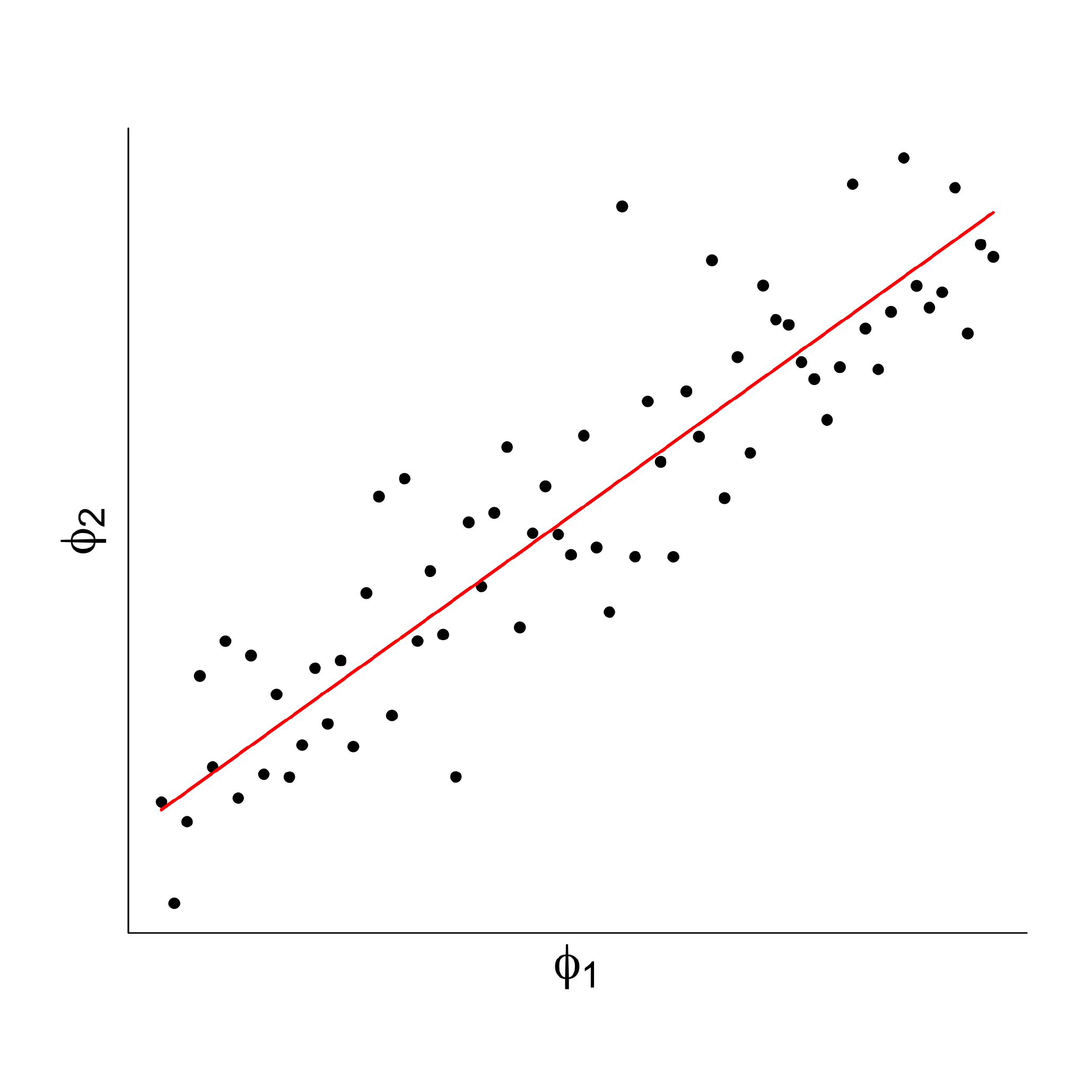}}
                \caption{ }
                \label{fig:dimensionReduction_a}
        \end{subfigure}
        \begin{subfigure}[b]{0.32\textwidth}
                \centerline{\includegraphics[width=\columnwidth]{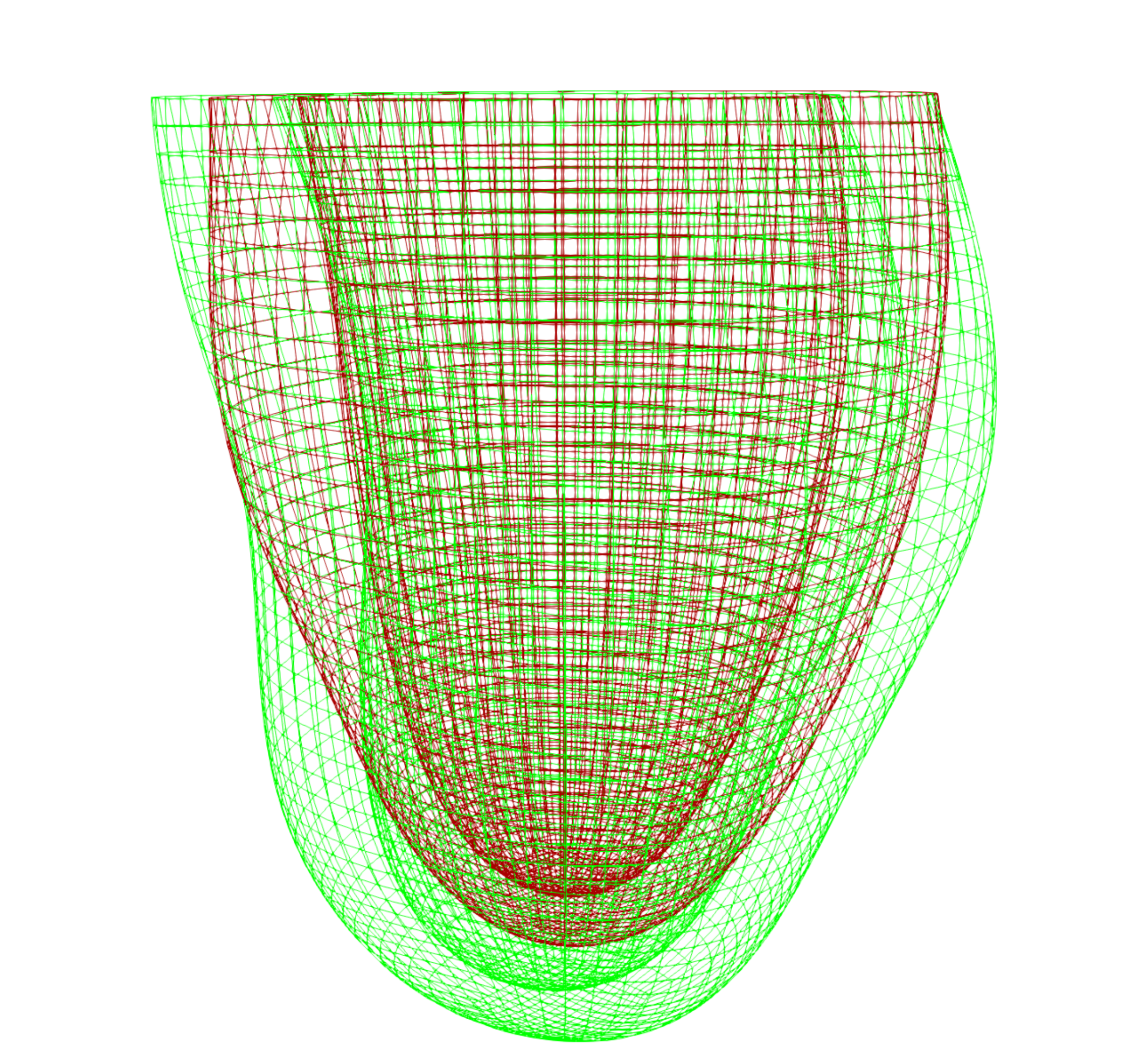}}
                \caption{ }
                \label{fig:dimensionReduction_b}
        \end{subfigure}
        \begin{subfigure}[b]{0.32\textwidth}
                \centerline{\includegraphics[width=\columnwidth]{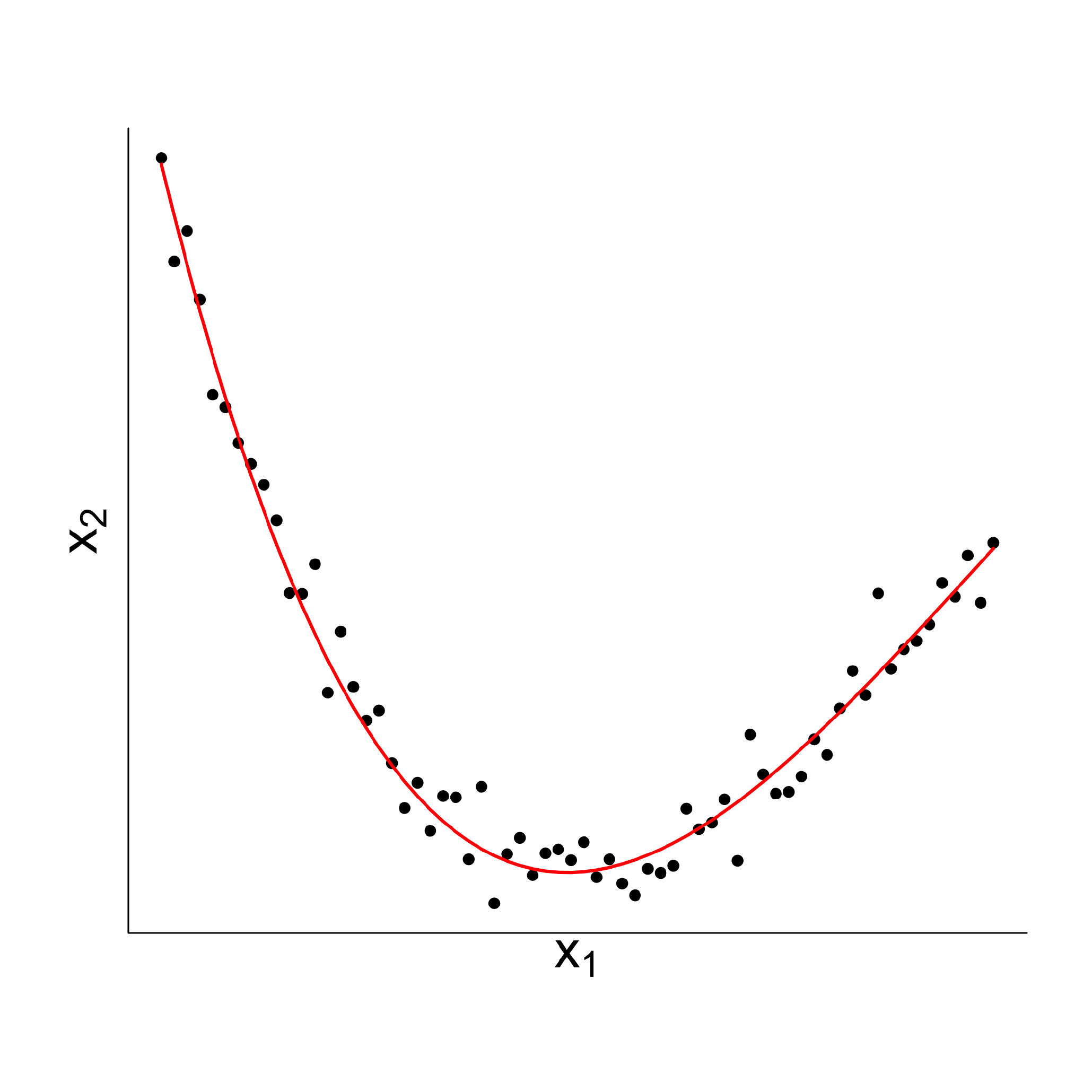}}
                \caption{ }                
                \label{fig:dimensionReduction_c}
        \end{subfigure}
        \caption{Illustration of dimension reduction for the representation of the left ventricle (LV). (a) Illustration of PCA. A set of LV geometries extracted from a set of patients forms a cloud of vectors in a high-dimensional vector space (here reduced to 2 for visual representation). PCA provides a set of linear orthogonal subspaces along the directions of maximum variance (here only one, the leading component, is shown). (b) A variation along the principal component can be mapped back into the high-dimensional vector space to show the corresponding changes of the LV geometry (here indicated by different colour shadings). (c) PCA is a linear technique and hence suboptimal if the LV geometries from the patient population are grouped along a non-linear submanifold. }
				\label{fig:dimensionReduction}
\end{figure}

A straightforward approach for achieving this low-dimensional representation is principle component analysis (PCA), illustrated in Figure~\ref{fig:dimensionReduction_a}, where the high-dimensional LV geometries are mapped onto a low-dimensional space that captures the maximum variation in the population. A variation along the PCA directions can be mapped back into the high-dimensional LV geometry space to illustrate typical organ deformations, as illustrated in Figure~\ref{fig:dimensionReduction_b}. However, while fast and easy to implement, the limitation of PCA is its restriction to linear subspaces. If the LV geometries extracted from the patient population are grouped along a non-linear submanifold in the high-dimensional LV geometry space, as illustrated in Figure~\ref{fig:dimensionReduction_c}, PCA is suboptimal. A variety of non-linear extensions of and alternatives to PCA have been proposed in the machine learning and computational statistics literature. The most straightforward extension is kernel PCA \citep{KPCA}, which conceptually maps the data non-linearly into a high-dimensional vector space and makes use of Mercer's theorem, whereby the scalar product in this high-dimensional space is equivalent to a kernel in the original data space and therefore never has to be computed explicitly. Alternative non-linear dimension reduction methods to be explored are generative topographic maps \citep{GTM}, self-organising maps \citep{SOM}, and 
variational auto-encoding neural networks \citep{AutoVarBayes}.



\section*{Acknowledgement}

\noindent
This work was funded by the British Heart Foundation, grant numbers PG/14/64/31043 and RE/18/6134217, and by the UK Engineering and Physical Sciences Research Council (EPSRC), grant number EP/N014642/1, as part of the SofTMech project. Benn Macdonald is supported by The Biometrika Trust, Fellowship number B0003. Umberto No\`{e} was supported by a Biometrika Scholarship.
Alan Lazarus is partially funded by a grant from GlaxoSmithKline plc.

\bibliographystyle{elsarticle-harv} 
\bibliography{umberto_references,References,zx1,tdh,BibliographyUN}

\end{document}